\documentstyle[12pt,epsf,epsfig]{article}
\newcount\timecount
\newcount\hours \newcount\minutes  \newcount\temp \newcount\pmhours

\hours = \time
\divide\hours by 60
\temp = \hours
\multiply\temp by 60
\minutes = \time
\advance\minutes by -\temp
\def\hour{\the\hours}
\def\minute{\ifnum\minutes<10 0\the\minutes
            \else\the\minutes\fi}
\def\clock{
\ifnum\hours=0 12:\minute\ AM
\else\ifnum\hours<12 \hour:\minute\ AM
      \else\ifnum\hours=12 12:\minute\ PM
            \else\ifnum\hours>12
                 \pmhours=\hours
                 \advance\pmhours by -12
                 \the\pmhours:\minute\ PM
                 \fi
            \fi
      \fi
\fi
}

\def\monthname{\relax\ifcase\month 0/\or January\or February\or
   March\or April\or May\or June\or July\or August\or September\or
   October\or November\or December\else\number\month/\fi}

\def\bold#1{\setbox0=\hbox{$#1$}%
     \kern-.025em\copy0\kern-\wd0
     \kern.05em\copy0\kern-\wd0
     \kern-.025em\raise.0433em\box0 }

\def\ga{\mathrel{\raise.3ex\hbox{$>$\kern-.75em\lower1ex\hbox{$\sim$}}}}
\def\la{\mathrel{\raise.3ex\hbox{$<$\kern-.75em\lower1ex\hbox{$\sim$}}}}
\def\gev{{\rm \, Ge\kern-0.125em V}}
\def\tev{{\rm \, Te\kern-0.125em V}}
\def\beq{\begin{equation}}
\def\eeq{\end{equation}}

\def\ohsq{\Omega_{\chi} h^2}

\def\m12{m_{1\!/2}}

\def\ga{\mathrel{\raise.3ex\hbox{$>$\kern-.75em\lower1ex\hbox{$\sim$}}}}
\def\la{\mathrel{\raise.3ex\hbox{$<$\kern-.75em\lower1ex\hbox{$\sim$}}}}
\def\gyr{{\rm \, G\kern-0.125em yr}}
\def\gev{{\rm \, Ge\kern-0.125em V}}
\def\tev{{\rm \, Te\kern-0.125em V}}
\def\beq{\begin{equation}}
\def\eeq{\end{equation}}

\def\stop{\tilde t}

\def\m12{m_{1\!/2}}

\def\gappeq{\mathrel{\rlap {\raise.5ex\hbox{$>$}}
{\lower.5ex\hbox{$\sim$}}}}

\def\lappeq{\mathrel{\rlap{\raise.5ex\hbox{$<$}}
{\lower.5ex\hbox{$\sim$}}}}

\def\Toprel#1\over#2{\mathrel{\mathop{#2}\limits^{#1}}}

\begin{document}
\begin{titlepage}
\pagestyle{empty}
\baselineskip=21pt
\rightline{hep-ph/0202110}
\rightline{CERN--TH/2002-027}
\rightline{UMN--TH--2042/02}
\rightline{TPI--MINN--02/02}
\vskip 1in
\begin{center}
{\large{\bf Constraining Supersymmetry}}
\end{center}
\begin{center}
\vskip 0.2in
{{\bf John Ellis}$^1$, {\bf Keith
A.~Olive}$^{2}$ and {\bf Yudi Santoso}$^{2}$}\\
\vskip 0.1in
{\it
$^1${TH Division, CERN, Geneva, Switzerland}\\
$^2${Theoretical Physics Institute,
University of Minnesota, Minneapolis, MN 55455, USA}}\\
\vskip 0.2in
{\bf Abstract}
\end{center}
\baselineskip=18pt \noindent

%%%%%%%%%%%%%%%%%%%%%%%%%%%%%%%%%%%%%%%%%%%%%%%%%%%%%%%%%%%%%%%%%%%%%

We review constraints on the minimal supersymmetric extension of the
Standard Model (MSSM) coming from direct searches at accelerators such as
LEP, indirect measurements such as $b \to s \gamma$ decay and the
anomalous magnetic moment of the muon. The recently corrected sign of pole
light-by-light scattering contributions to the latter is taken into
account. We combine these constraints with those due to the cosmological
density of stable supersymmetric relic particles. The possible indications
on the supersymmetric mass scale provided by fine-tuning arguments are
reviewed critically. We discuss briefly the prospects for future
accelerator searches for supersymmetry.

%%%%%%%%%%%%%%%%%%%%%%%%%%%%%%%%%%%%%%%%%%%%%%%%%%%%%%%%%%%%%%%%%%%%%
\vskip 0.5in

{\it Invited Contribution to the New Journal of Physics Focus Issue on
Supersymmetry}

\vfill
\leftline{CERN--TH/2002-027}
%%%%%%%%%%%%%%%%%%%%%%%%%%%%%%%%%%%%%%%%%%%%%%%%%%%%%%%%%%%%%%%%%%%%%%%%%%%%%%
\end{titlepage}

\section{Introduction}

The avoidance of fine tuning has long been the primary motivation for
supersymmetry at the TeV scale~\cite{hierarchy}. This issue is normally 
formulated in
connection with the hierarchy problem: why/how is $m_W \ll m_P$, or
equivalently why is $G_F \sim 1/m^2_W \gg G_N = 1/m^2_P$, or equivalently
why does the Coulomb potential in an atom dominate over the Newton
potential, $e^2 \gg G_N m_p m_e \sim (m/m_P)^2$, where $m_{p,e}$ are the
proton and electron masses? One might think naively that it would be
sufficient to set $m_W \ll m_P$ by hand. However, radiative corrections
tend to destroy this hierarchy. For example, one-loop diagrams generate
\beq
\delta m^2_W = {\cal O}\left({\alpha\over\pi}\right)~\Lambda^2 \gg m^2_W
\label{four}
\eeq
where $\Lambda$ is a cut-off representing the appearance of new physics, and
the
inequality in (\ref{four}) applies if $\Lambda\sim 10^3$ TeV, and even 
more so if  $\Lambda \sim m_{GUT} \sim
10^{16}$
GeV or $ \sim m_P \sim 10^{19}$ GeV. If the radiative corrections to a 
physical
quantity
are much larger than its measured values, obtaining the latter requires
strong
cancellations, which in general require fine tuning of the bare input
parameters.
However, the necessary cancellations are natural in supersymmetry, where 
one has
equal
numbers of bosons $B$ and fermions $F$ with equal couplings, so that
(\ref{four})
is replaced by
\beq
\delta m^2_W = {\cal O}\left({\alpha\over\pi}\right)~\vert m^2_B - 
m^2_F\vert~.
\label{five}
\eeq
The residual radiative correction is naturally small if
\beq
\vert m^2_B - m^2_F\vert \lappeq 1~{\rm TeV}^2
\label{six}
\eeq
As we shall see later, cosmology also favours the mass range (\ref{six}) 
for the lightest supersymmetric particle (LSP), if it is stable. In this 
case, the LSP would be an excellent candidate for astrophysical dark
matter~\cite{EHNOS}. In the following the LSP is assumed to be a 
neutralino $\chi$, i.e., a mixture of the $\tilde\gamma, \tilde H$ and $\tilde 
Z$.

The minimal supersymmetric extension of the Standard Model (MSSM) has the
same gauge interactions as the Standard Model, and similar Yukawa
couplings. A key difference is the necessity of two Higgs doublets, in
order to give masses to all the quarks and leptons, and to cancel triangle
anomalies. This duplication is important for phenomenology: it means that
there are five physical Higgs bosons, two charged $H^\pm$ and three
neutral $h, H, A$. Their quartic self-interactions are determined by the
gauge interactions, solving the vacuum instability problem mentioned above
and limiting the possible mass of the lightest neutral Higgs boson.
However, the doubling of the Higgs multiplets introduces two new
parameters: $\tan\beta$, the ratio of Higgs vacuum expectation values and
$\mu$, a parameter mixing the two Higgs doublets.

The MSSM predicts that there should appear a Higgs boson weighing $\lappeq
130$~GeV. Thus, fans of supersymmetry have been encouraged by the fact
that the precision electroweak data favour~\cite{LEPEWWG} a relatively 
light Higgs boson
with $m_H \simeq$ 115 GeV, just above the exclusion unit provided by
direct searches at LEP. They were even more encouraged by the possible
sighting during the last days of LEP of a Higgs boson, with a preferred
mass of 115.6 GeV~\cite{LEPHiggs}. If this were to be confirmed, it would
suggest that the Standard Model breaks down at some relatively low energy
$\lappeq 10^3$ TeV~\cite{ER}. Above this scale the effective Higgs
potential of the Standard Model becomes unstable as the quartic Higgs
self-coupling is driven negative by radiative corrections due to the
relatively heavy top quark~\cite{tdown}. This is not necessarily a
disaster, and it is possible that the present electroweak vacuum might be
metastable, provided that its lifetime is longer than the age of the
Universe~\cite{Setal}. However, we would surely feel more secure if such
an instability could be avoided by introducing suitable new physics below
$10^3$ TeV. However, any new physics must be finely tuned, or the
potential blows up instead~\cite{ER}, and this fine tuning also occurs 
naturally in
supersymmetry. However, this argument is logically distinct from the
previous hierarchy argument. There supersymmetry was motivated by the
control of quadratic divergences, and here by the absence of logarithmic
divergences.

Another experimental hint in favour of supersymmetry is provided by the
LEP measurements of the gauge couplings, that are in very good agreement
with supersymmetric GUTs~\cite{susyGUTs}, again if sparticles weigh $\sim
1$~TeV. This argument does not provide a strong constraint on the 
supersymmetry-breaking scale, particularly because there may be 
important threshold effects near the GUT scale, but it does favour 
qualitatively models with accessible sparticles.

A crucial ingredient in the MSSM is the soft supersymmetry breaking, in
the form of scalar masses $m_0$, gaugino masses $m_{1/2}$ and trilinear
couplings $A$~\cite{DG}. These are presumed to be inputs from physics at 
some high-energy scale, e.g., from some supergravity or superstring theory,
which then evolve down to lower energy scale according to well-known
renormalization-group equations. In the case of the Higgs multiplets, this
renormalization can drive the effective mass-squared negative, triggering
electroweak symmetry breaking~\cite{renn}. It is often assumed that 
the $m_0$ are universal at the input scale~\footnote{Universality between the
squarks and sleptons of different generations is motivated by upper limits
on flavour-changing neutral interactions~\cite{FCNC}, but universality 
between the soft masses of the $L, E^c, Q^c, D^c$ and $U^c$ is not so well
motivated.}, as are the $m_{1/2}$ and $A$ parameters. In this case the
free parameters are
\beq
m_0, m_{1/2}, A \quad {\rm and}\quad \tan\beta~,
\label{eight}
\eeq
with $\mu$ being determined by the electroweak vacuum conditions, up to a 
sign. We refer to this scenario as the constrained MSSM (CMSSM).

\section{Constraints on the MSSM}

\begin{figure}
\vskip 0.5in
\vspace*{-0.75in}
%\hspace*{-.70in}
\begin{minipage}{8in}
\epsfig{file=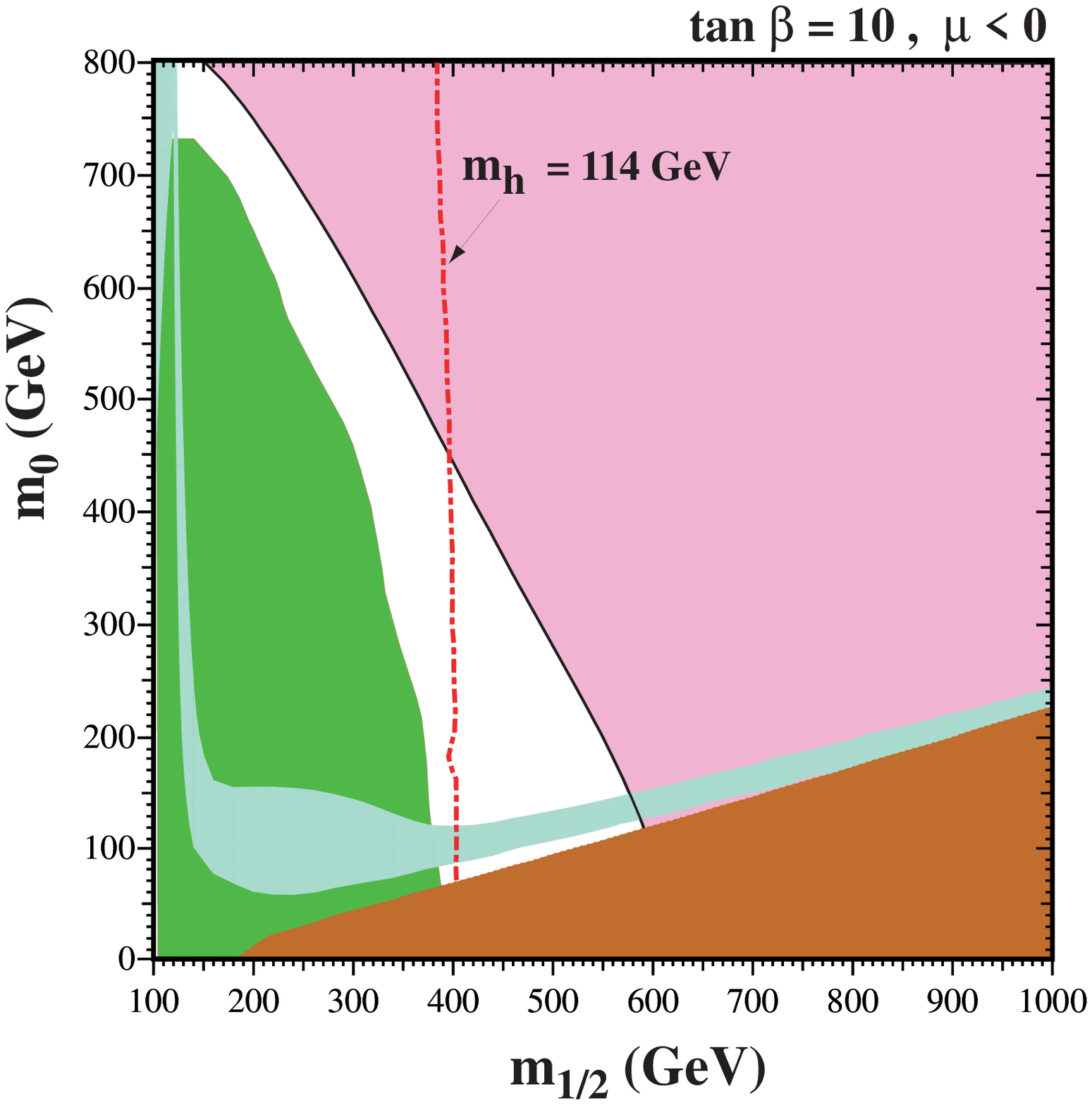,height=3.3in}
\hspace*{-0.17in}
\epsfig{file=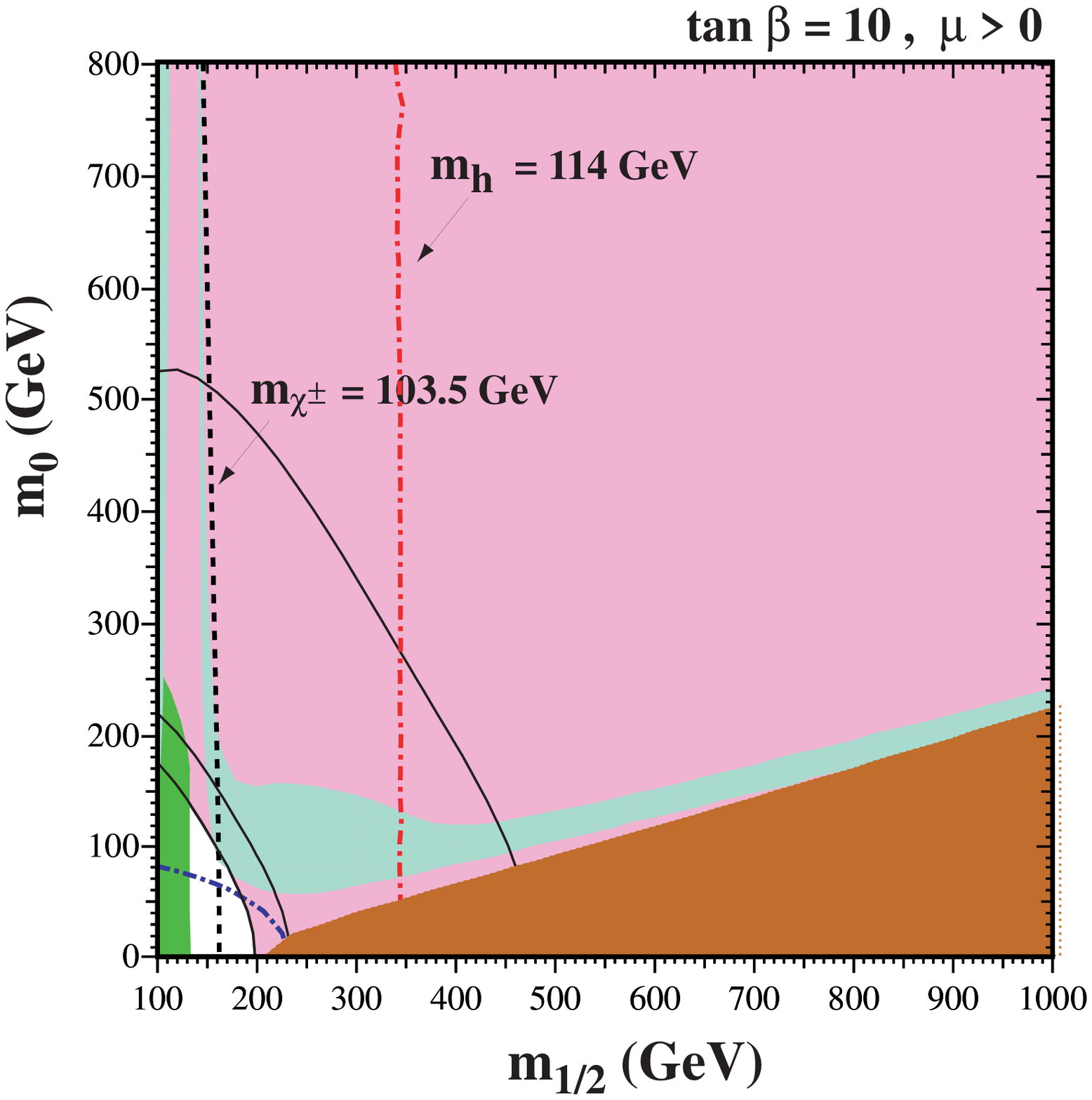,height=3.3in} \hfill
\end{minipage}
%\vspace*{-3in}
%\hspace*{-.70in}
\begin{minipage}{8in}
%\hskip -1.40in
%\vskip -.75in
\epsfig{file=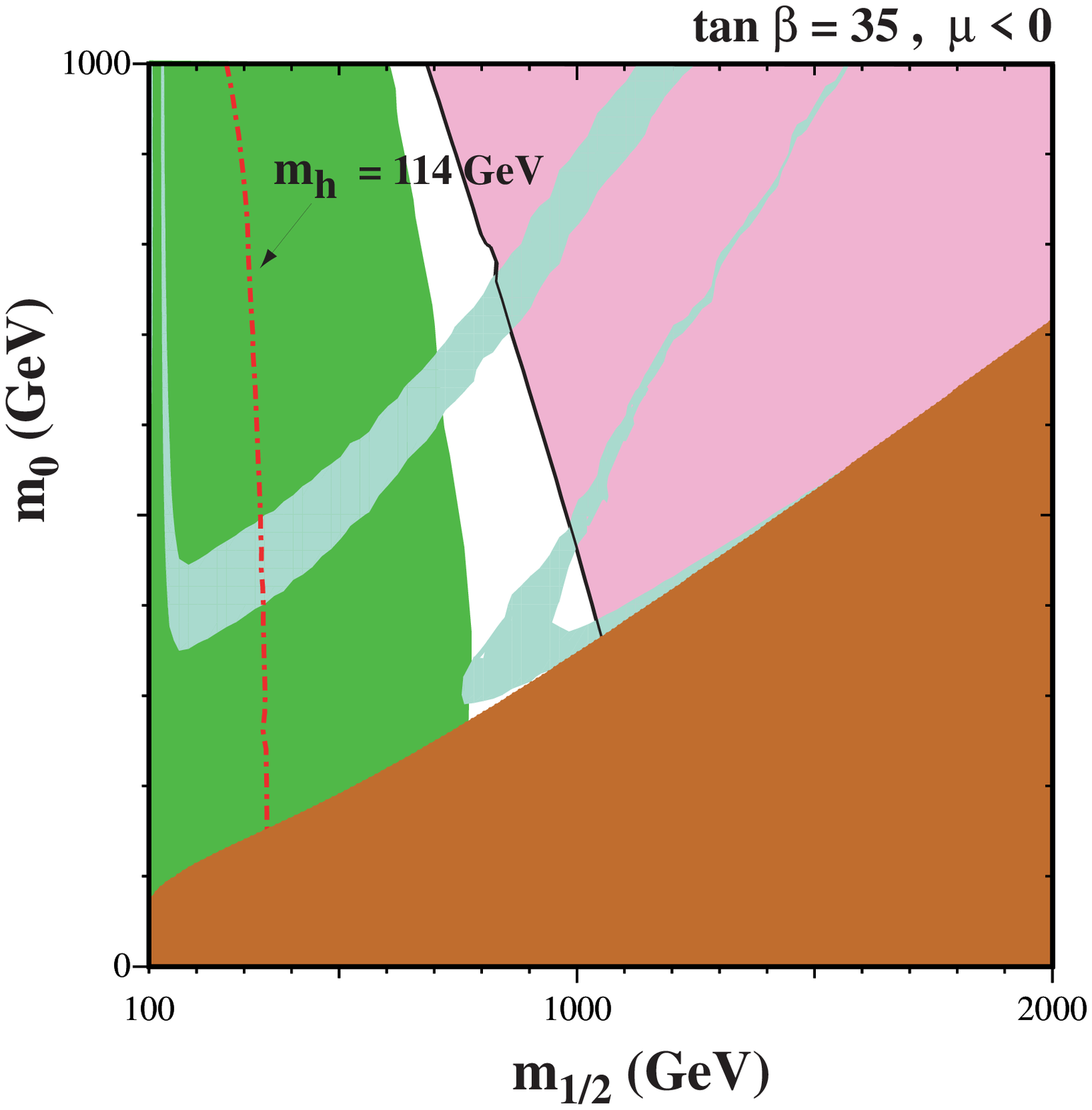,height=3.3in}
\hspace*{-0.2in}
\epsfig{file=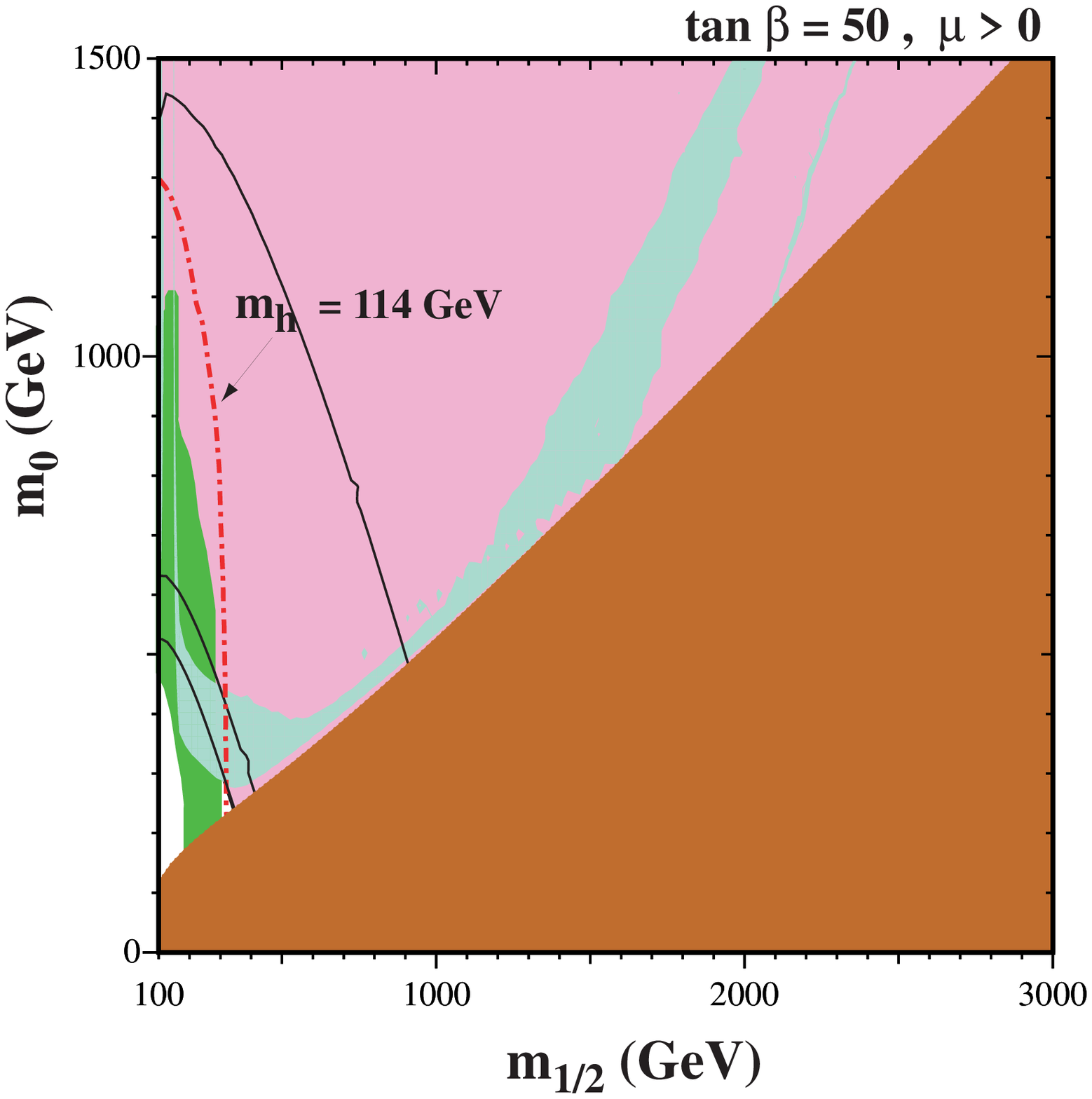,height=3.3in} \hfill
\end{minipage}
%\vskip 2.5in
\caption{
{\it Compilations of phenomenological constraints on the CMSSM for
(a) $\tan \beta = 10, \mu < 0$,  (b) $\tan \beta = 10, \mu > 0$, (c)
$\tan \beta = 35, \mu < 0$ and (d)  $\tan \beta = 50, \mu > 0$, assuming
$A_0 = 0, m_t = 175$~GeV and $m_b(m_b)^{\overline {MS}}_{SM} = 4.25$~GeV
\cite{EFGOSi}.  The near-vertical lines are the LEP limits
$m_{\chi^\pm} = 103.5$~GeV (dashed and black)~\cite{LEPsusy}, shown in
(b) only, and
$m_h = 114.1$~GeV (dotted and red)~\cite{LEPHiggs}. 
Also, in the lower left corner of (b), we show the $m_{\tilde e} = 99$
GeV contour \protect\cite{LEPSUSYWG_0101}.  In the dark (brick red)
shaded regions, the LSP is the charged
${\tilde
\tau}_1$, so this region is excluded. The light (turquoise) shaded areas
are the cosmologically preferred regions with
\protect\mbox{$0.1\leq\ohsq\leq 0.3$}~\cite{EFGOSi}. The medium (dark
green) shaded regions that are most prominent in panels (a) and (c) are
excluded by $b \to s \gamma$~\cite{bsg}. The shaded (pink) regions in the 
upper right regions delineate the $\pm 2 \, \sigma$ ranges of $g_\mu -
2$. For $\mu > 0$, the $\pm 1 \, \sigma$ contours are also shown as solid 
black lines.
}}
\label{fig:CMSSM}
\end{figure}

Important experimental constraints on the MSSM parameter space are
provided by direct searches at LEP and the Tevatron collider, as
seen in Fig.~\ref{fig:CMSSM} in the case of the CMSSM. One of these is the
limit $m_{\chi^\pm} \gappeq$ 103.5 GeV provided by chargino searches at 
LEP
\cite{LEPsusy},
where the third significant figure depends on other CMSSM parameters. LEP
has also provided lower limits on slepton masses, of which the strongest
is $m_{\tilde e}\gappeq$ 99 GeV \cite{LEPSUSYWG_0101}, again depending
only sightly on the other CMSSM parameters, as long as $m_{\tilde e} -
m_\chi \gappeq$ 10 GeV. The most important constraints on the $u, d, s,
c, b$ squarks and gluinos are provided by the Tevatron collider: for
equal masses
$m_{\tilde q} = m_{\tilde g} \gappeq$ 300 GeV. In the case of the $\tilde
t$, LEP provides the most stringent limit when $m_{\tilde t} - m_\chi$ is
small, and the Tevatron for larger $m_{\tilde t} - m_\chi$~\cite{LEPsusy}.

Another important constraint is provided by the LEP lower limit on the
Higgs mass: $m_H > $ 114.1 GeV \cite{LEPHiggs}. This holds in the
Standard Model, for the lightest Higgs boson $h$ in the general MSSM for
$\tan\beta
\lappeq 8$, and almost always in the CMSSM for all $\tan\beta$, at least
as long as CP is conserved~\footnote{The lower bound on the lightest MSSM
Higgs boson may be relaxed significantly if CP violation feeds into the
MSSM Higgs sector~\cite{CEPW}.}. Since $m_h$ is sensitive to sparticle
masses, particularly $m_{\tilde t}$, via loop corrections:
\beq
\delta m^2_h \propto {m^4_t\over m^2_W}~\ln\left({m^2_{\tilde t}\over
m^2_t}\right)~ + \ldots
\label{nine}
\eeq
the Higgs limit also imposes important constraints on the CMSSM parameters,
principally $m_{1/2}$ as seen in Fig.~\ref{fig:CMSSM}. The constraints are 
evaluated using {\tt FeynHiggs}~\cite{FeynHiggs}, which is estimated to 
have a residual uncertainty of a couple of GeV in $m_h$.

Also shown in Fig.~\ref{fig:CMSSM} is the constraint imposed by
measurements of $b\rightarrow s\gamma$~\cite{bsg}. These agree with the
Standard Model, and therefore provide bounds on MSSM particles, 
such as the chargino and charged Higgs
masses, in particular. Typically, the $b\rightarrow s\gamma$
constraint is more important for $\mu < 0$, as seen in
Fig.~\ref{fig:CMSSM}a and c, but it is also relevant for $\mu > 0$, 
particularly when $\tan\beta$ is large as seen in Fig.~\ref{fig:CMSSM}d.

The final experimental constraint we consider is that due to the
measurement of the anamolous magnetic moment of the muon.  The BNL E821
experiment reported last year a new measurement of
$a_\mu\equiv {1\over 2} (g_\mu -2)$ which deviated by 2.6 standard
deviations from the best Standard Model prediction available at that
time~\cite{BNL}. The largest contribution to the errors in the comparison
with theory was thought to be the statistical error of the experiment,
which will soon be significantly reduced, as many more data have already
been recorded. However, it has recently been realized that the sign of
the most important pseudoscalar-meson pole part of the light-by-light 
scattering contribution
\cite{lightbylight} to the Standard Model prediction should be reversed,
which reduces the apparent experimental discrepancy to about 1.6 standard
deviations. The next-largest error is thought to be that due to
strong-interaction uncertainties in the Standard Model prediction, for
which recent estimates converge to about 7$\times 10^{-10}$~\cite{DHSNTY}.

As many authors have pointed out~\cite{susygmu}, a discrepancy between
theory and the BNL experiment could well be explained by supersymmetry. As
seen in Fig.~\ref{fig:CMSSM}, this is particularly easy if $\mu > 0$.
With the change in sign of the meson-pole contributions to
light-by-light scattering, good consistency is also possible for $\mu <
0$ so long as either $m_{1/2}$ or $m_0$ are taken sufficiently large.
We show in Fig.~\ref{fig:CMSSM} as medium (pink) shaded the new $2 \,
\sigma$  allowed region: $-6 < \delta a_\mu \times 10^{10} <  58 $.

The new regions preferred by the $g-2$ experimental data shown in 
Fig.~\ref{fig:CMSSM} differ considerably from the older
ones~\cite{susygmu} which were based on the range $11 < \delta a_\mu
\times 10^{10} <  75 $.  First of all, the older bound completely 
excluded $\mu < 0$ at the $2 \, \sigma$ level.  As one can see this is no
longer treu.  $\mu < 0 $ is allowed so long as either (or both) $m_{1/2}$
and $m_0$ are large. Thus for $\mu < 0$, one is forced into either
the $\chi-{\tilde \tau}$ coannihilation region or the funnel region
produced by the s-channel annihilatn via the heavy Higgses $H$ and $A$.
Second, whereas the older limits produced definite upper bounds on the
sparticle masses (which were accept with delight to future collider
builders), the new bounds which are consistent with $a_mu = 0$, 
allow arbitrarily high sparticle masses.  Now only the very low mass
corner of the $m_{1/2}-m_0$ plane is excluded.

Fig.~\ref{fig:CMSSM} also displays the regions where the supersymmetric 
relic density $\rho_\chi = \Omega_\chi \rho_{critical}$ falls within the 
preferred range
\beq
0.1 < \Omega_\chi h^2 < 0.3
\label{ten}
\eeq
The upper limit is rigorous, since astrophysics and cosmology tell us that
the total matter density $\Omega_m \lappeq 0.4$, and the Hubble expansion
rate $h \sim 1/\sqrt{2}$ to within about 10 \% (in units of 100 km/s/Mpc). On
the other hand, the lower limit in (\ref{ten}) is optional, since there
could be other important contributions to the overall matter density.

As is seen in Fig.~\ref{fig:CMSSM}, there are generic regions of the CMSSM
parameter space where the relic density falls within the preferred range
(\ref{ten}). What goes into the calculation of the relic density? It is
controlled by the annihilation cross section~\cite{EHNOS}:
\beq
\rho_\chi = m_\chi n_\chi \, , \quad n_\chi \sim {1\over
\sigma_{ann}(\chi\chi\rightarrow\ldots)}\, ,
\label{eleven}
\eeq
where the typical annihilation cross section $\sigma_{ann} \sim 
1/m_\chi^2$.
For this reason, the relic density typically increases with the relic
mass, and this combined with the upper bound in (\ref{ten}) then leads to
the common expectation that $m_\chi \lappeq$ O(200) GeV. 

\begin{figure}
\vskip 0.5in
\vspace*{-0.75in}
%\hspace*{-.70in}
\begin{minipage}{8in}
\epsfig{file=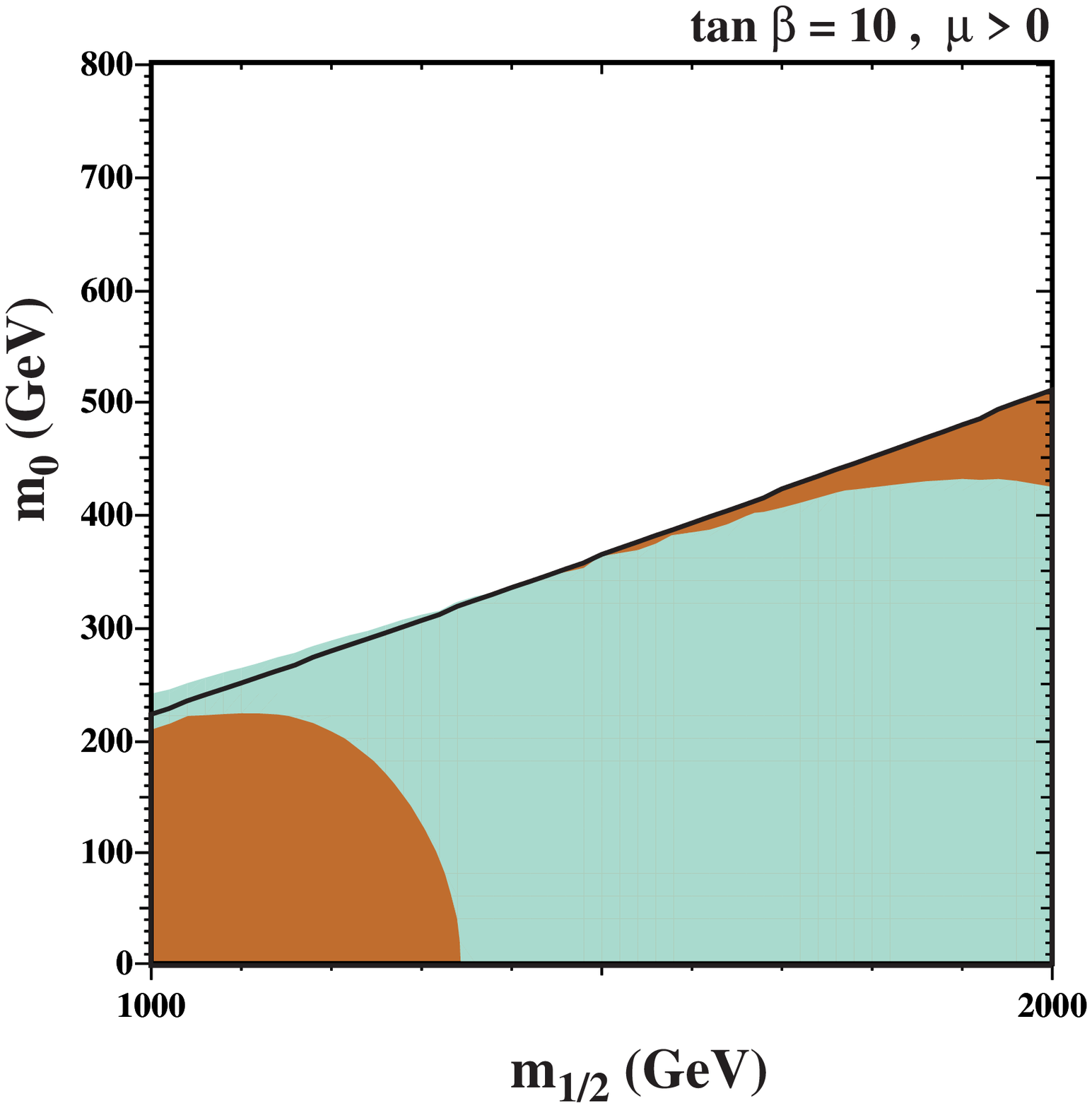,height=3.3in}
\hspace*{-0.17in}
\epsfig{file=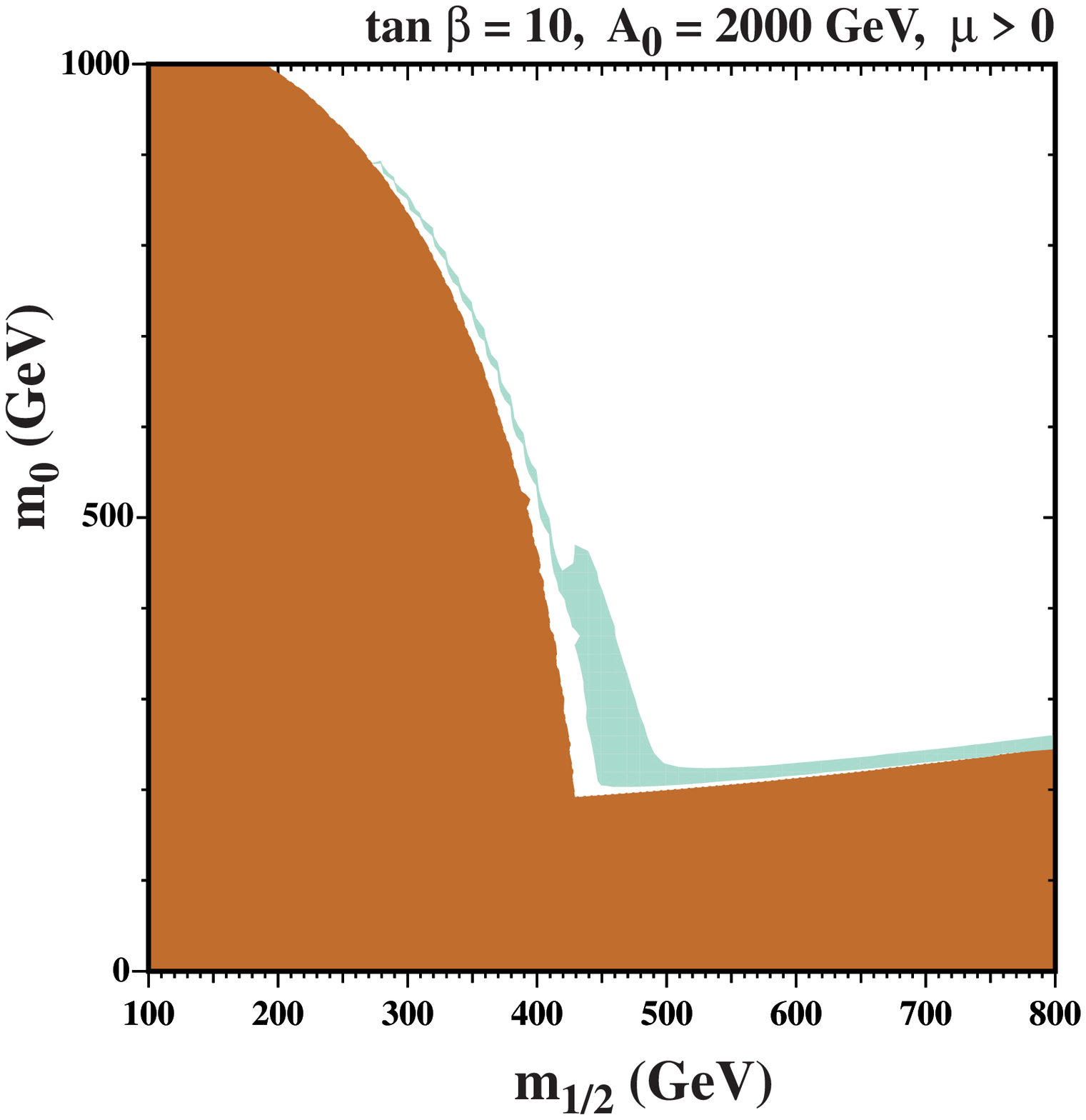,height=3.3in} \hfill
\end{minipage}
\caption[]{\it (a) The large-$m_{1/2}$ `tail' of the $\chi - {\tilde 
\tau_1}$ 
coannihilation region 
for $\tan \beta = 10$, $A = 0$ and $\mu < 0$~\cite{ourcoann}, superimposed 
on the disallowed dark (brick red) shaded region where $m_{\tilde
\tau_1} < m_\chi$, and (b) the $\chi - {\tilde t_1}$ coannihilation region
for $\tan \beta = 10$, $A = 2000$~GeV and $\mu > 0$~\cite{EOS}, exhibiting 
a large-$m_0$ `tail'.}
\label{fig:coann}
\end{figure}

\begin{figure}
%\vspace*{-0.75in}
\hspace*{-.40in}
\begin{minipage}{8in}
\epsfig{file=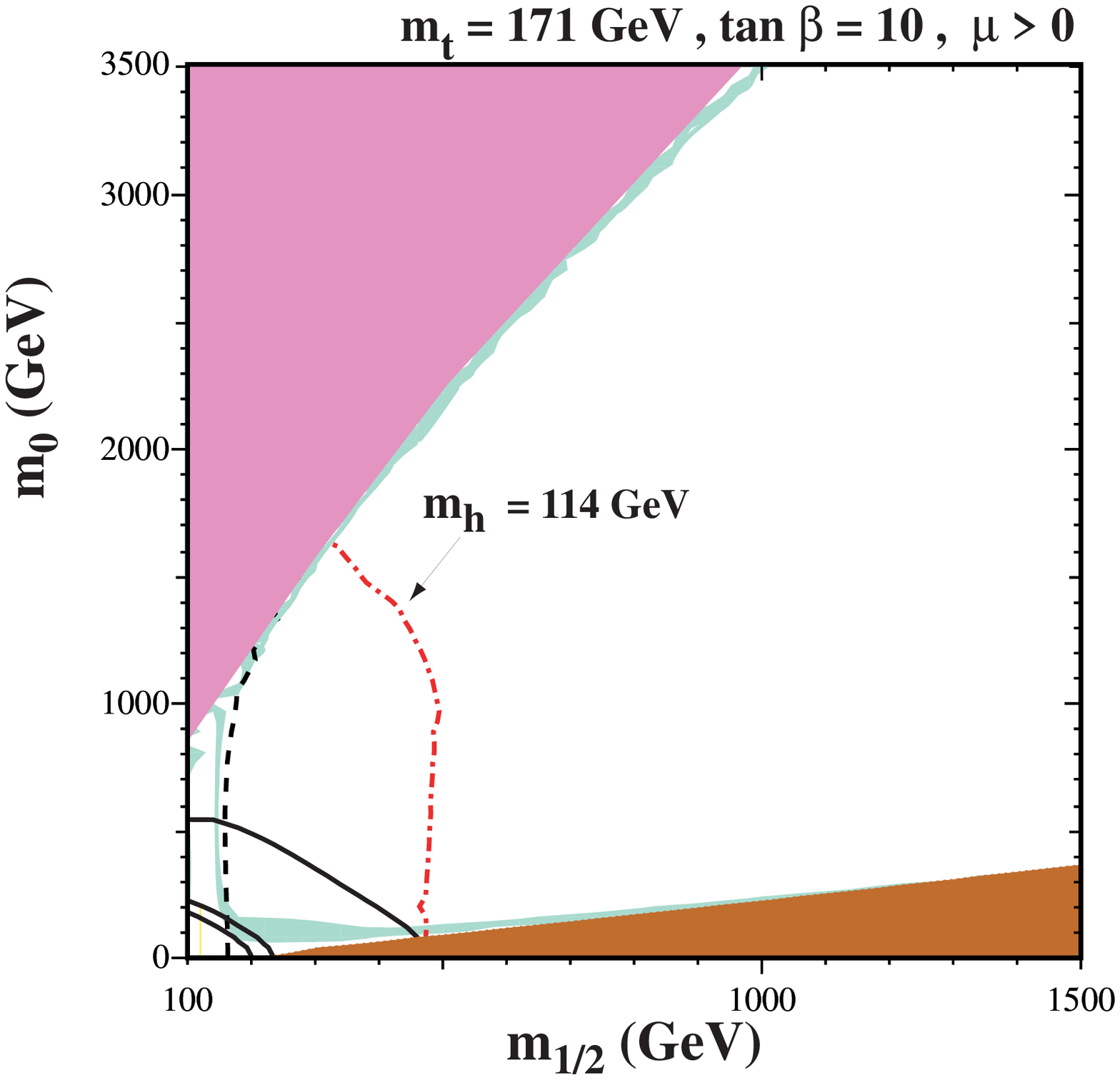,height=3.3in}
\hspace*{-0.17in}
\epsfig{file=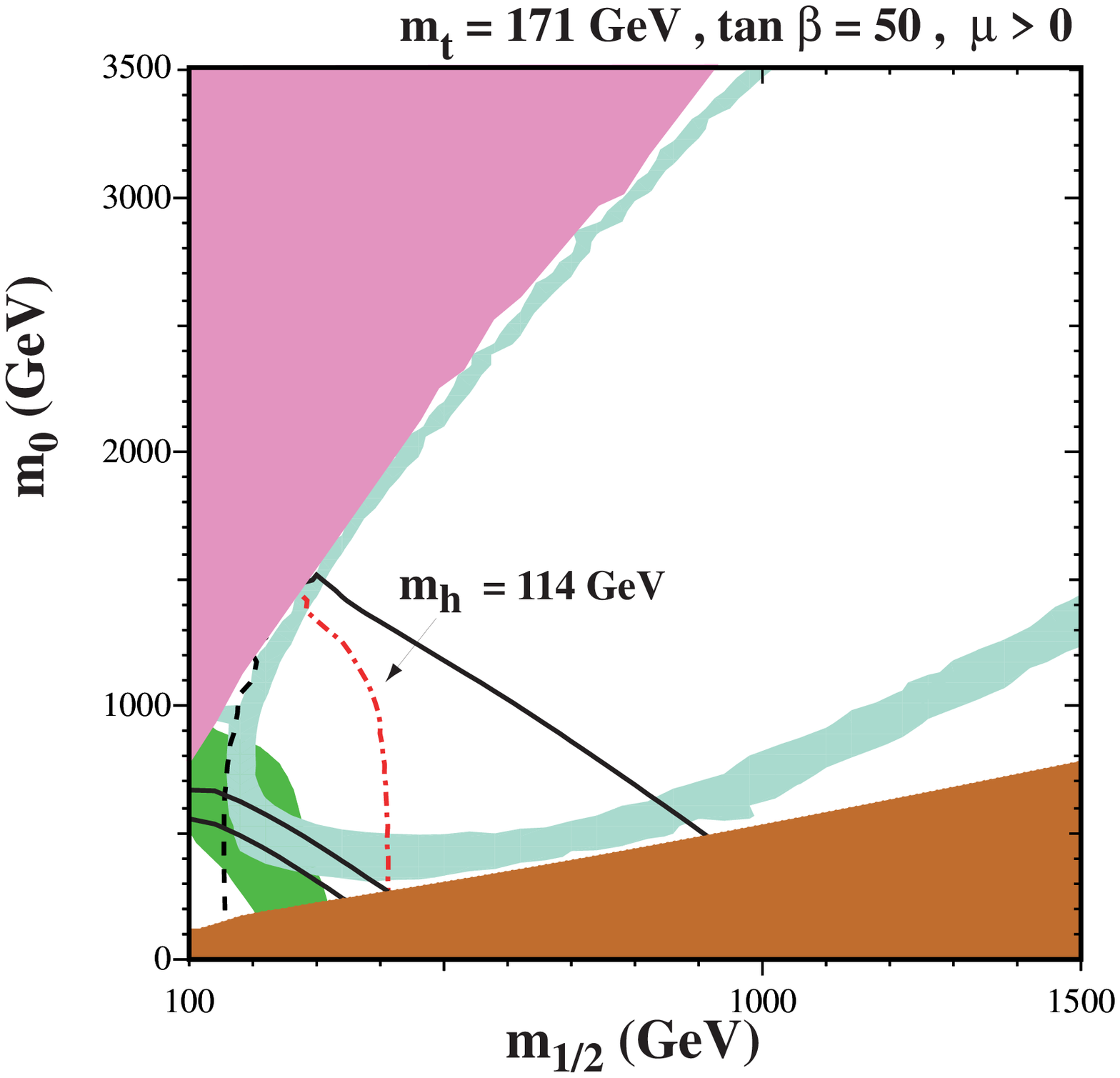,height=3.3in} \hfill
\end{minipage}
\caption[]{\it An expanded view of the $m_{1/2} - m_0$ parameter plane
showing the focus-point regions \protect\cite{focus} at large $m_0$ for 
(a) $tan \beta
= 10$, and (b) $\tan \beta = 50$. In the shaded (mauve) region in the 
upper
left corner, there are no solutions with proper electroweak symmetry 
breaking, so these are
excluded in the CMSSM.  Note that we have chosen $m_t = 171$ GeV, in 
which case the focus-point region is at lower $m_0$ than when $m_t = 175$ 
GeV, as assumed in the other figures. The position of this region
is very sensitive to $m_t$. The black contours (both dashed and solid)
are as in Fig.~\protect\ref{fig:CMSSM}, the we do not shade the preferred
$g-2$ region. }
\label{fig:focus}
\end{figure}

However, there are various ways in which the generic upper bound on
$m_\chi$ can be increased along filaments in the $(m_{1/2},m_0)$ plane.
For example, if the next-to-lightest sparticle (NLSP) is not much heavier
than $\chi$: $\Delta m/m_\chi \lappeq 0.1$, the relic density may be
suppressed by coannihilation: $\sigma (\chi + $NLSP$ \rightarrow \ldots
)$~\cite{coann}.  In this way, the allowed CMSSM region may acquire a
`tail' extending to larger sparticle masses. An example of this 
possibility is the
case where the NLSP is the lighter stau: $\tilde\tau_1$ and
$m_{\tilde\tau_1} \sim m_\chi$, as seen in Figs.~\ref{fig:CMSSM}(a) and
(b) and extended to larger $m_{1/2}$ in
Fig.~\ref{fig:coann}(a)~\cite{ourcoann}. Another example is
coannihilation when the NLSP is the lighter stop
\cite{stopco}: $\tilde t_1$ and
$m_{\tilde t_1}
\sim m_\chi$, which may be important in the general MSSM or in the CMSSM
when
$A$ is large, as seen in Fig.~\ref{fig:coann}(b)~\cite{EOS}. In
the cases studied, the upper limit on $m_\chi$ is not affected by stop
coannihilation. Another mechanism for extending the allowed CMSSM region
to large
$m_\chi$ is rapid annihilation via a direct-channel pole when $m_\chi
\sim {1\over 2} m_{Higgs, Z}$~\cite{funnel,EFGOSi}. This may yield a
`funnel' extending to large $m_{1/2}$ and $m_0$ at large $\tan\beta$, as
seen in panels (c) and (d) of  Fig.~\ref{fig:CMSSM}~\cite{EFGOSi}. Yet
another allowed region at large
$m_{1/2}$ and $m_0$ is the `focus-point' region~\cite{focus}, which is
adjacent to the boundary of the region where electroweak symmetry breaking
is possible, as seen in Fig.~\ref{fig:focus}.

\section{Fine Tuning}

\begin{figure}
\vskip 0.5in
\vspace*{-0.75in}
\hspace*{-.20in}
\begin{minipage}{8in}
\epsfig{file=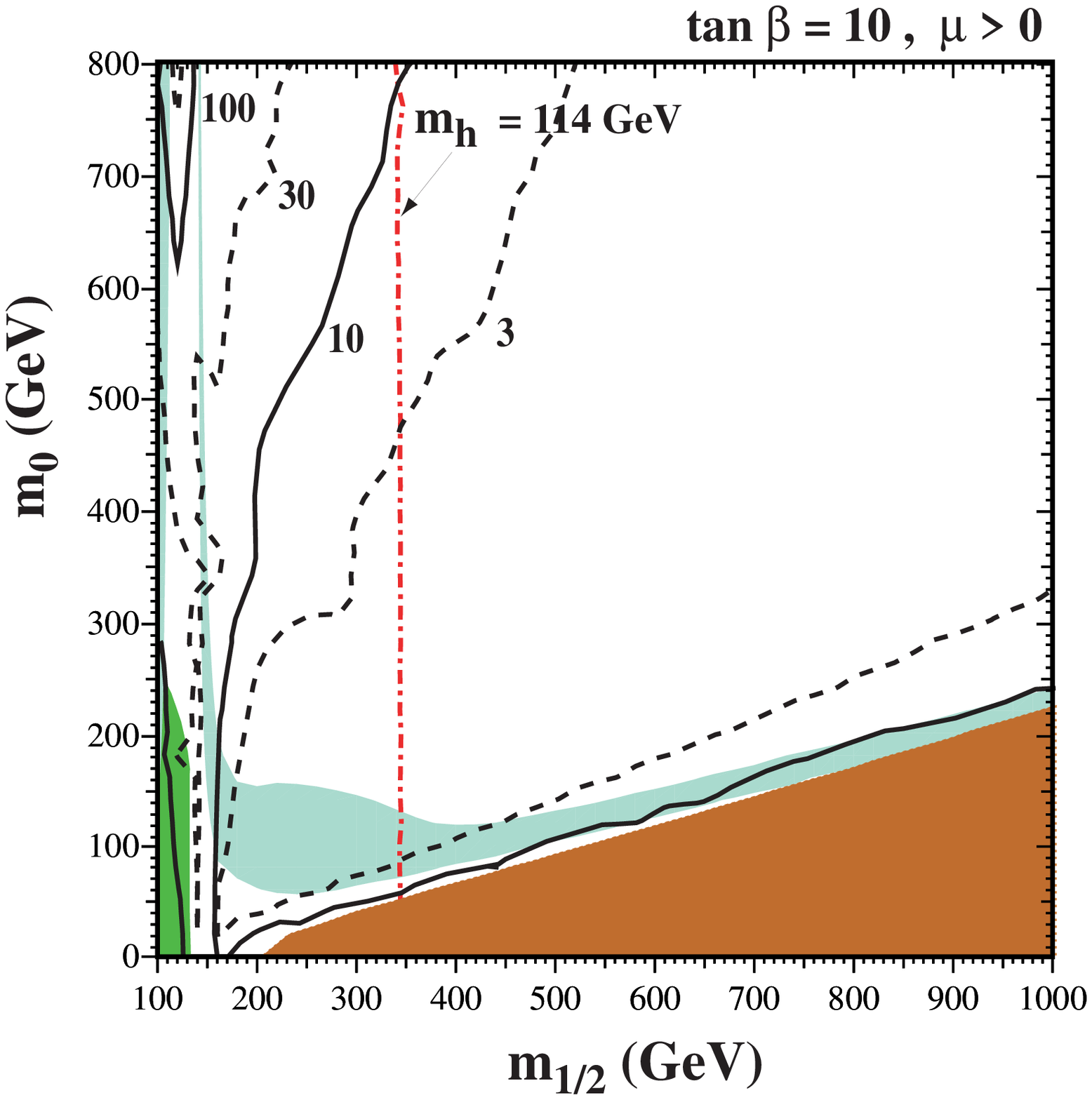,height=3.3in}
%\hspace*{-0.17in}
\epsfig{file=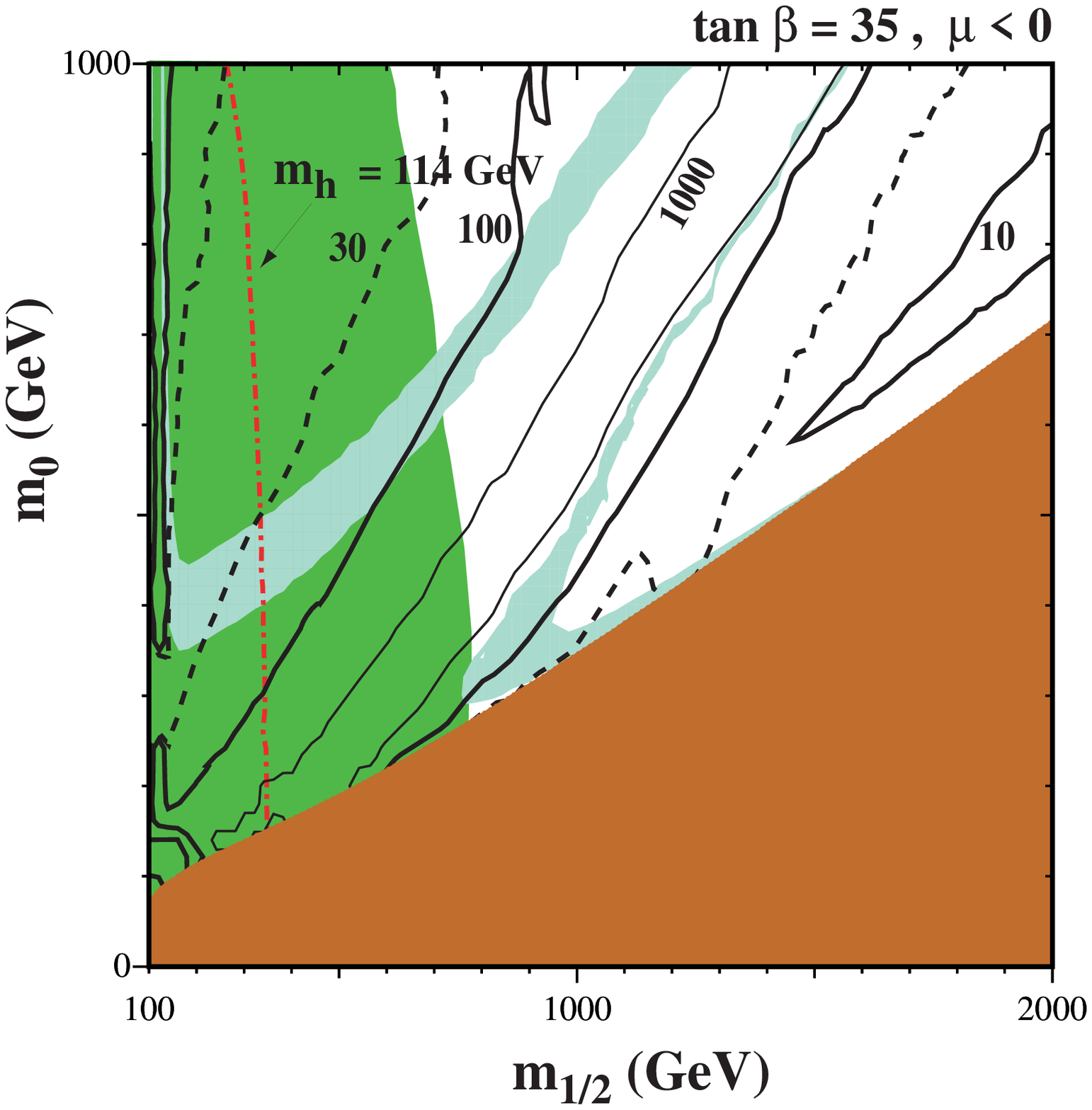,height=3.3in} \hfill
\end{minipage}
%\vspace*{-3in}
%\hspace*{-.70in}
\hspace*{-.20in}
\begin{minipage}{8in}
%\hskip -1.40in
%\vskip -.75in
\epsfig{file=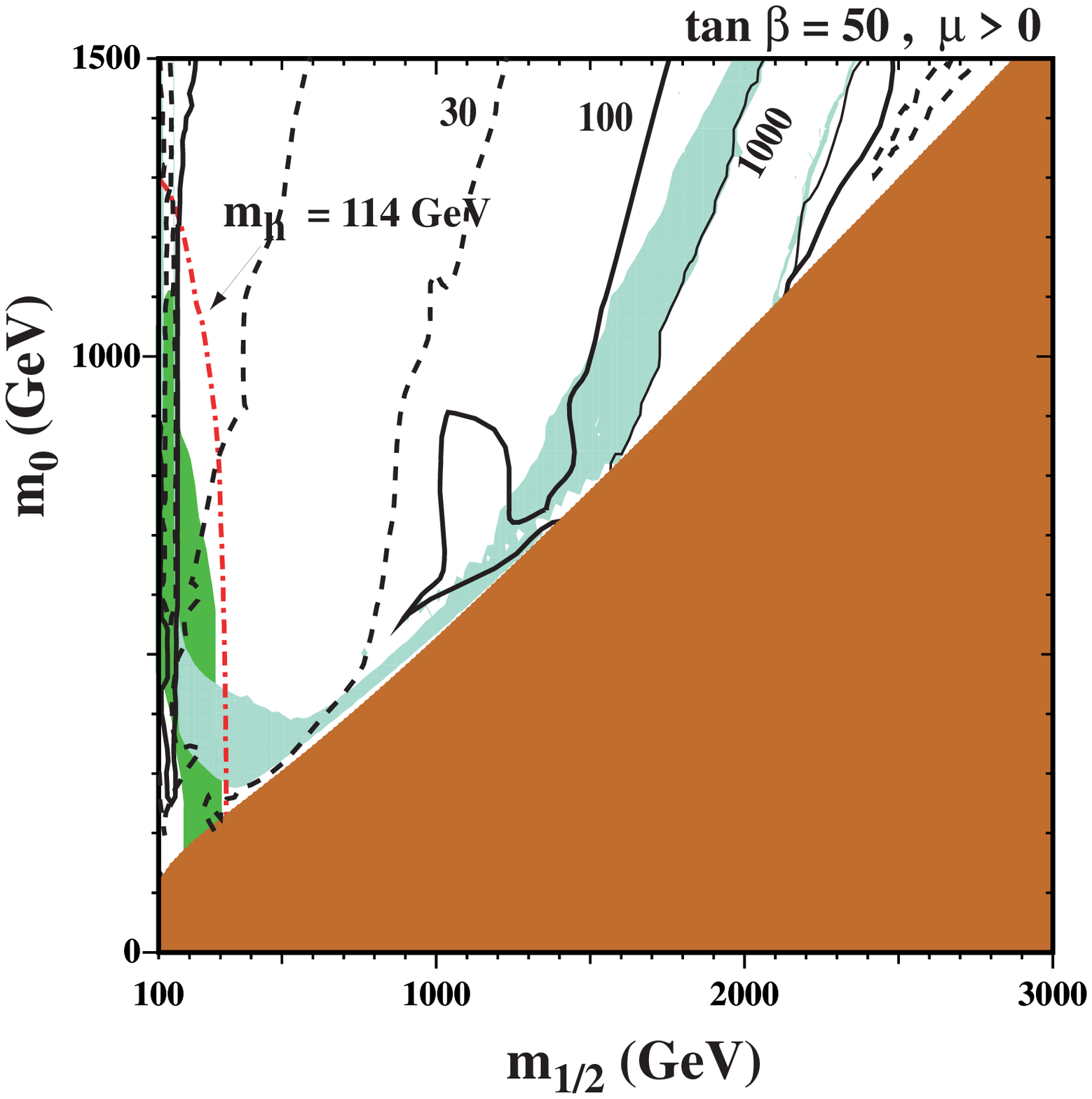,height=3.3in}
\hspace*{-0.10in}
\epsfig{file=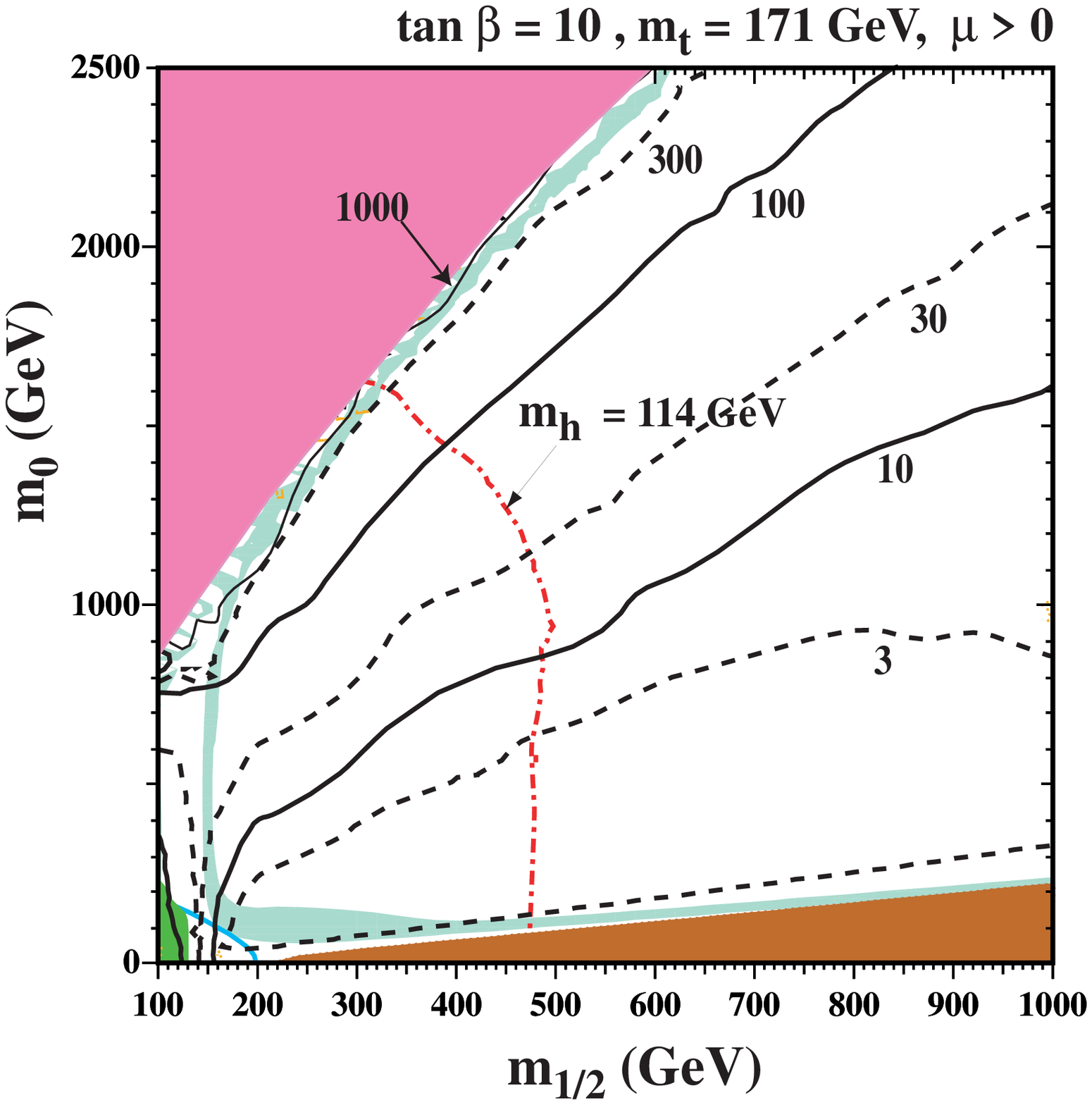,height=3.3in} \hfill
\end{minipage}
%\vskip 2.5in 
\caption{\label{fig:overall}
{\it Contours of the total sensitivity $\Delta^\Omega$ (\ref{twelve}) of 
the relic density in the
$(m_{1/2}, m_0)$ planes for (a) $\tan \beta = 10, \mu > 0, m_t =
175$~GeV, (b) $\tan \beta = 35, \mu < 0, m_t = 175$~GeV, (c)
$\tan \beta = 50, \mu > 0, m_t = 175$~GeV, and (d) $\tan \beta =
10, \mu > 0, m_t = 171$~GeV, all for $A_0 = 0$. The light (turquoise)
shaded areas are the cosmologically preferred regions with
\protect\mbox{$0.1\leq\ohsq\leq 0.3$}. In the dark (brick red) shaded
regions, the LSP is the charged ${\tilde \tau}_1$, so these regions are
excluded. In panel (d), the medium shaded (mauve) region is excluded by
the electroweak vacuum conditions. }}
\end{figure}

The filaments extending the preferred CMSSM parameter space are clearly
exceptional, in some sense, so it is important to understand the sensitivity
of the relic density to input parameters, unknown higher-order effects, 
etc. One proposal is the relic-density fine-tuning measure~\cite{EO}
\beq
\Delta^\Omega \equiv \sqrt{\sum_i ~~\left({\partial\ln (\Omega_\chi h^2)\over
\partial
\ln a_i}\right)^2 }
\label{twelve}
\eeq
where the sum runs over the input parameters, which might include
(relatively) poorly-known Standard Model quantities such as $m_t$ and
$m_b$, as well as the CMSSM parameters $m_0, m_{1/2}$, etc. As seen in
Fig.~\ref{fig:overall}, the sensitivity $\Delta^\Omega$ (\ref{twelve}) is 
relatively small
in the `bulk' region at low $m_{1/2}$, $m_0$, and $\tan\beta$. However, it
is somewhat higher in the $\chi - \tilde\tau_1$ coannihilation `tail', and
at large $\tan\beta$ in general. The sensitivity measure $\Delta^\Omega$
(\ref{twelve}) is particularly high in the rapid-annihilation `funnel' and
in the `focus-point' region. This explains why published relic-density
calculations may differ in these regions~\cite{otherOmega}, whereas they 
agree well when
$\Delta^\Omega$ is small: differences may arise because of small
differences in the treatments of the inputs.

\begin{figure}
\vskip 0.5in
\vspace*{-0.75in}
\hspace*{-.20in}
\begin{minipage}{8in}
\epsfig{file=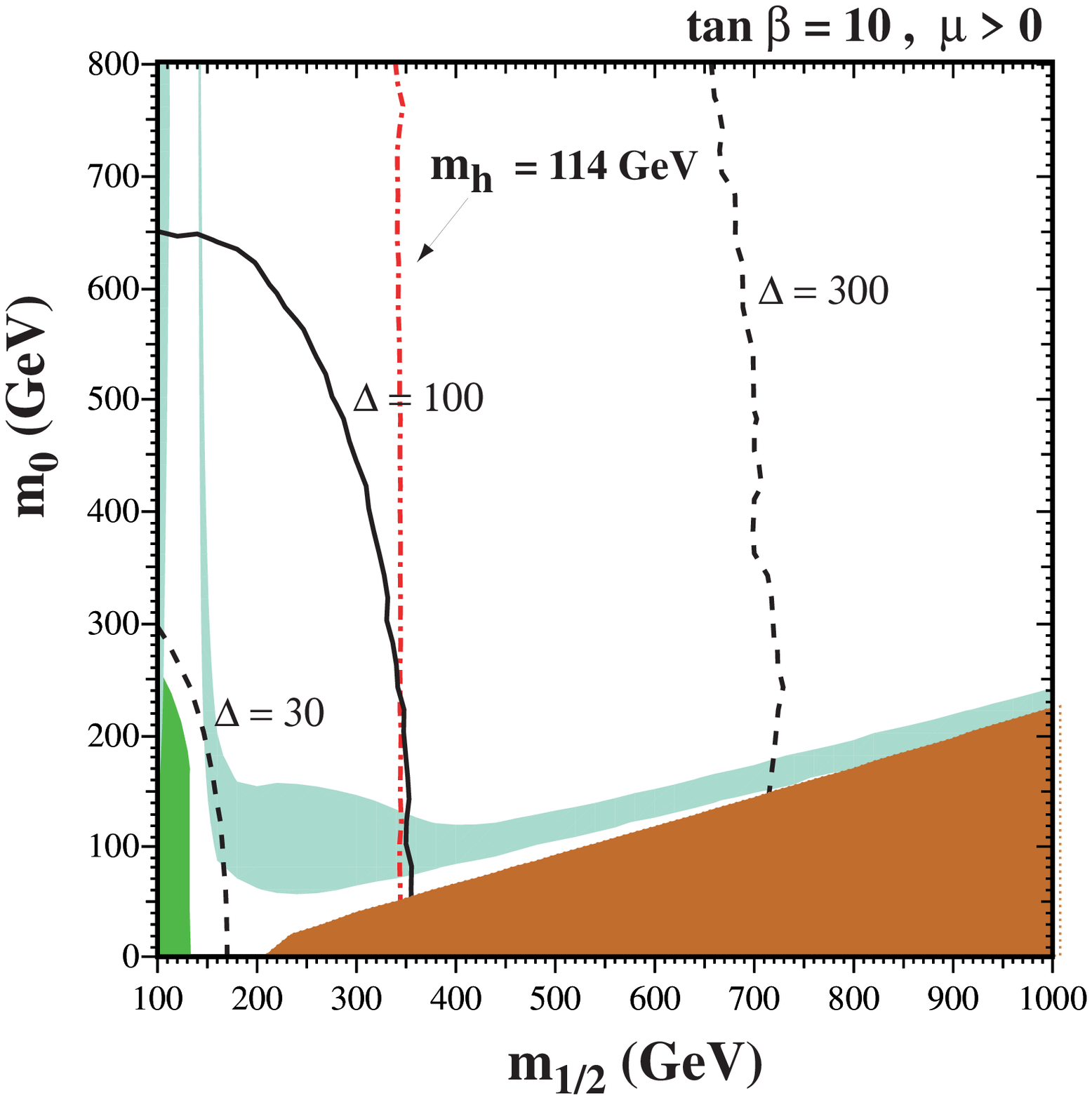,height=3.3in}
%\hspace*{-0.17in}
\epsfig{file=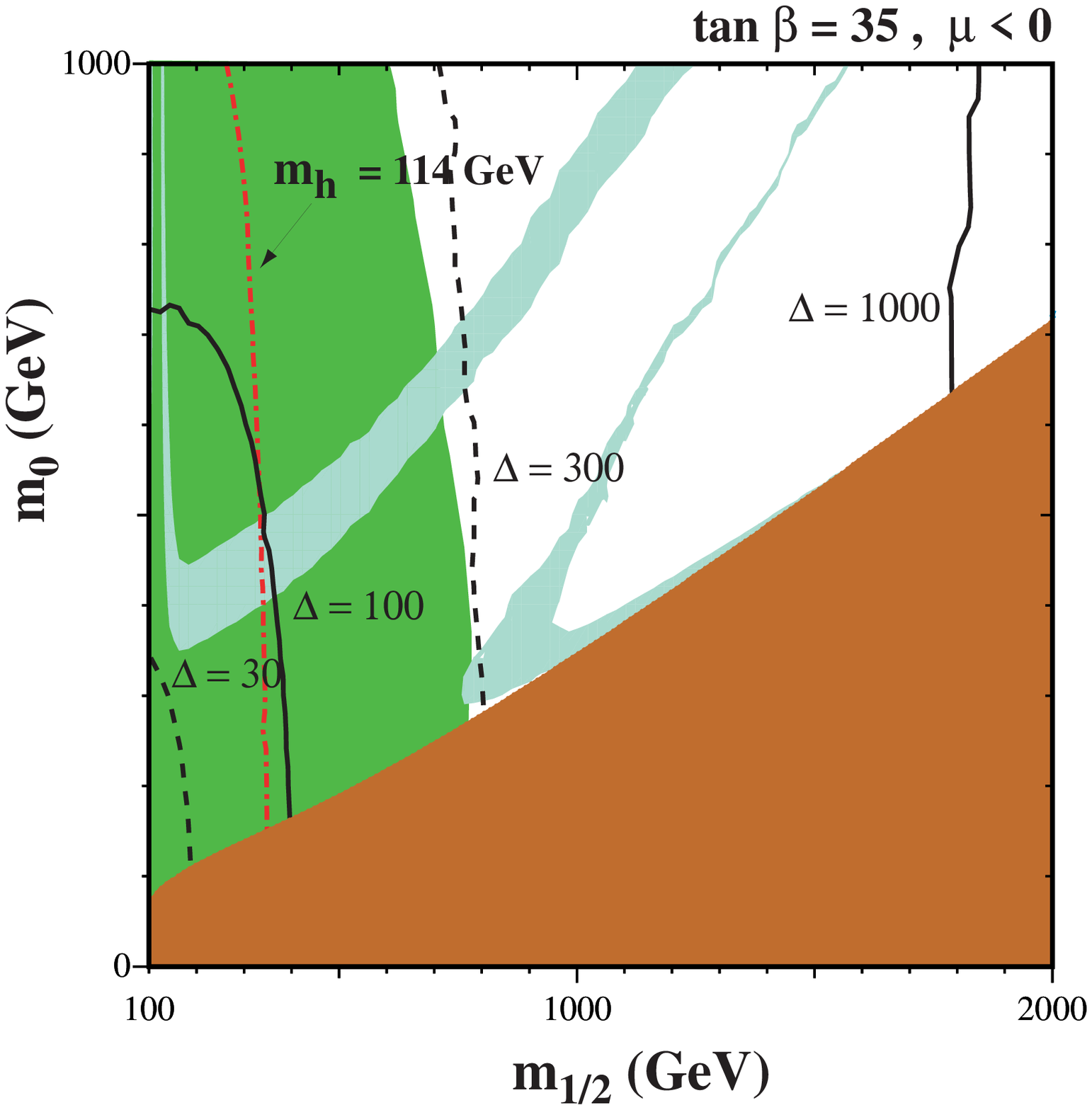,height=3.3in} \hfill
\end{minipage}
%\vspace*{-3in}
%\hspace*{-.70in}
\hspace*{-.20in}
\begin{minipage}{8in}
%\hskip -1.40in
%\vskip -.75in
\epsfig{file=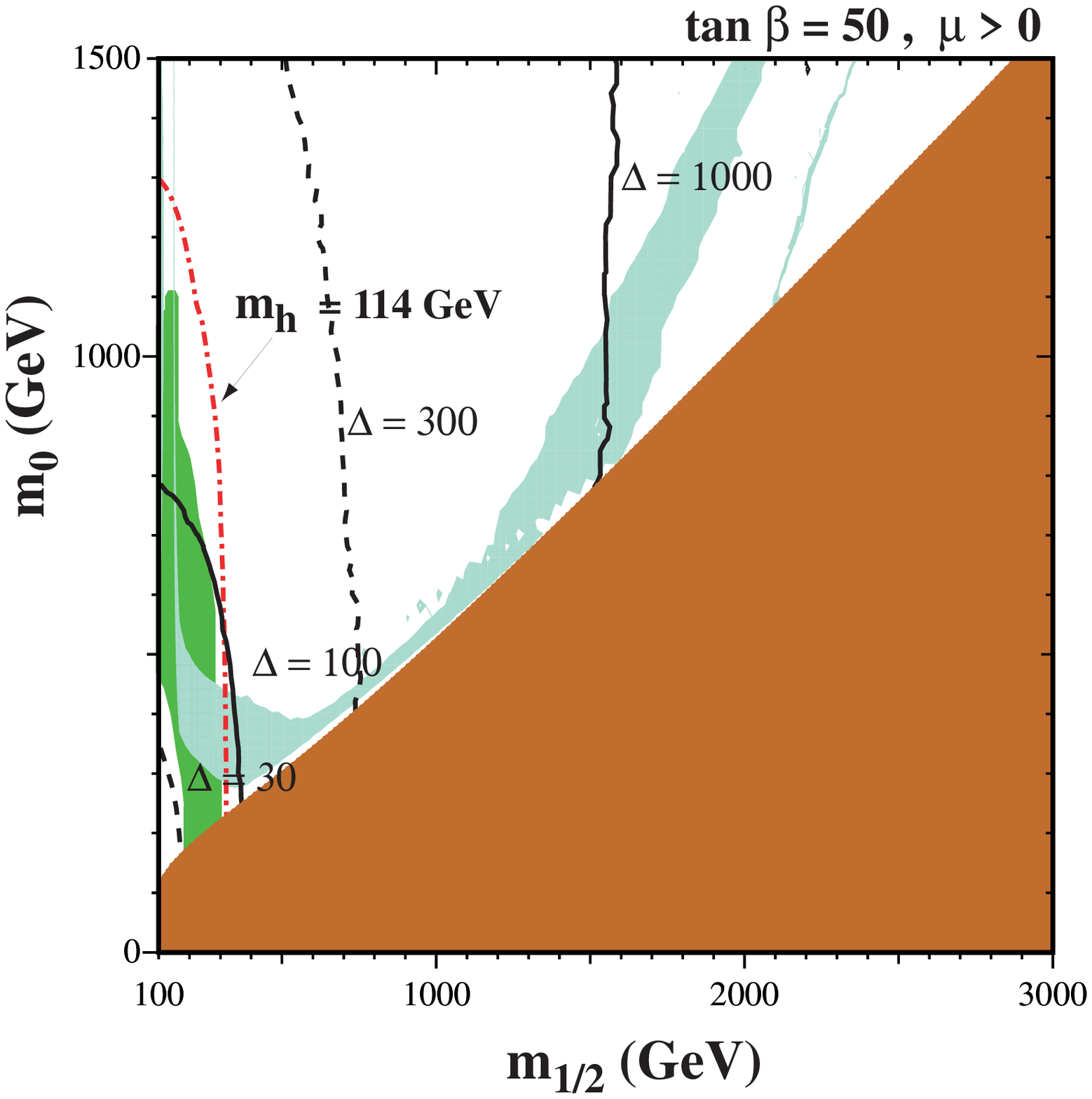,height=3.3in}
\hspace*{-0.20in}
\epsfig{file=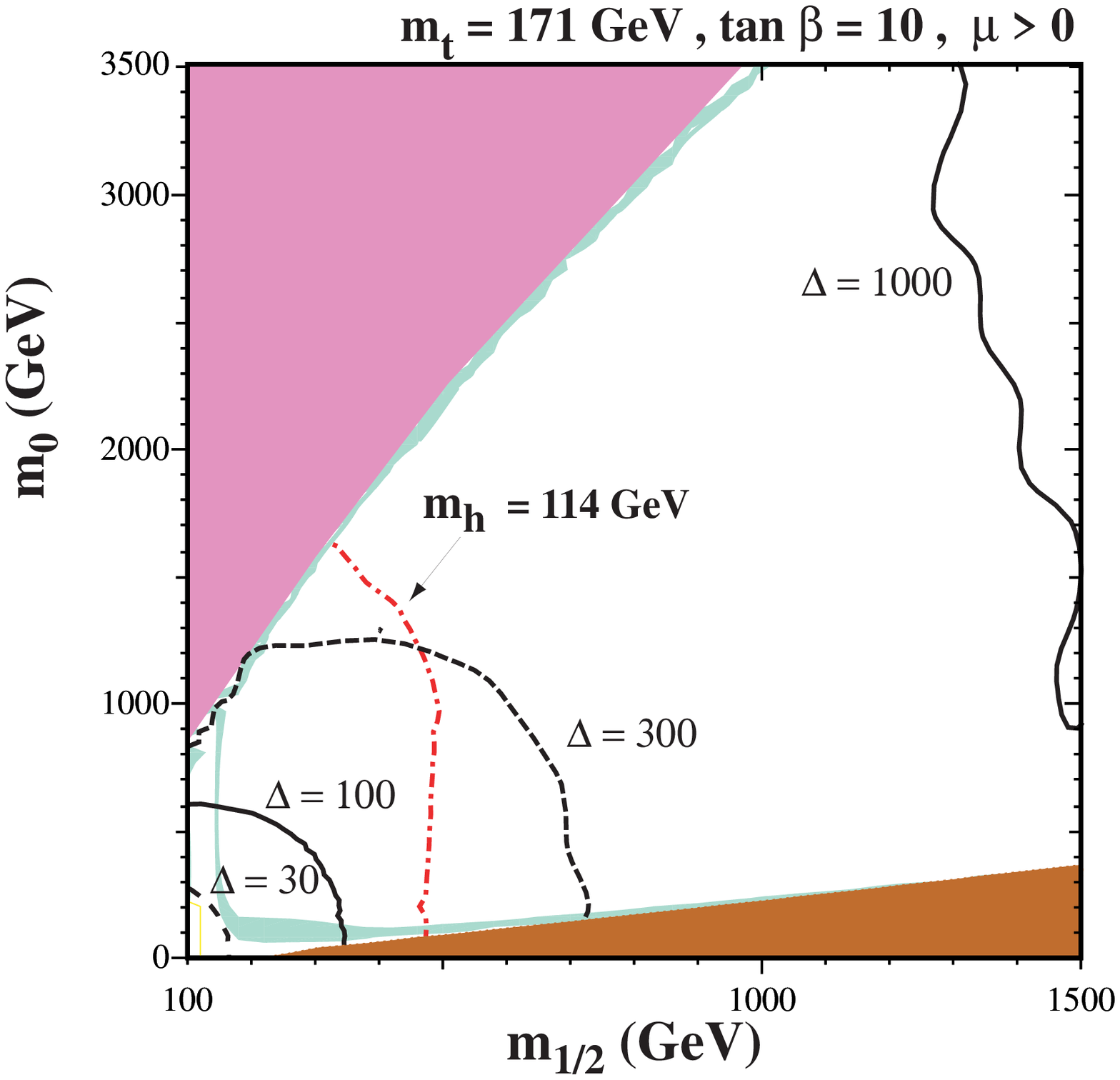,height=3.3in} \hfill
\end{minipage}
%\vskip 2.5in 
\caption{\label{fig:EWFT}
{\it 
Contours of the electroweak fine-tuning measure $\Delta$ 
(\ref{thirteen}) in the
$(m_{1/2}, m_0)$ planes for (a) $\tan \beta = 10, \mu > 0, m_t =
175$~GeV, (b) $\tan \beta = 35, \mu < 0, m_t = 175$~GeV, (c)
$\tan \beta = 50, \mu > 0, m_t = 175$~GeV, and (d) $\tan \beta =
10, \mu > 0, m_t = 171$~GeV, all for $A_0 = 0$. The light (turquoise)
shaded areas are the cosmologically preferred regions with
\protect\mbox{$0.1\leq\ohsq\leq 0.3$}. In the dark (brick red) shaded
regions, the LSP is the charged ${\tilde \tau}_1$, so this region is
excluded. In panel (d), the medium shaded (mauve) region is excluded by
the electroweak vacuum conditions. }}
\end{figure}  

\begin{figure}
\vskip 0.5in
\vspace*{-0.75in}
\hspace*{-.20in}
\begin{minipage}{8in}
\epsfig{file=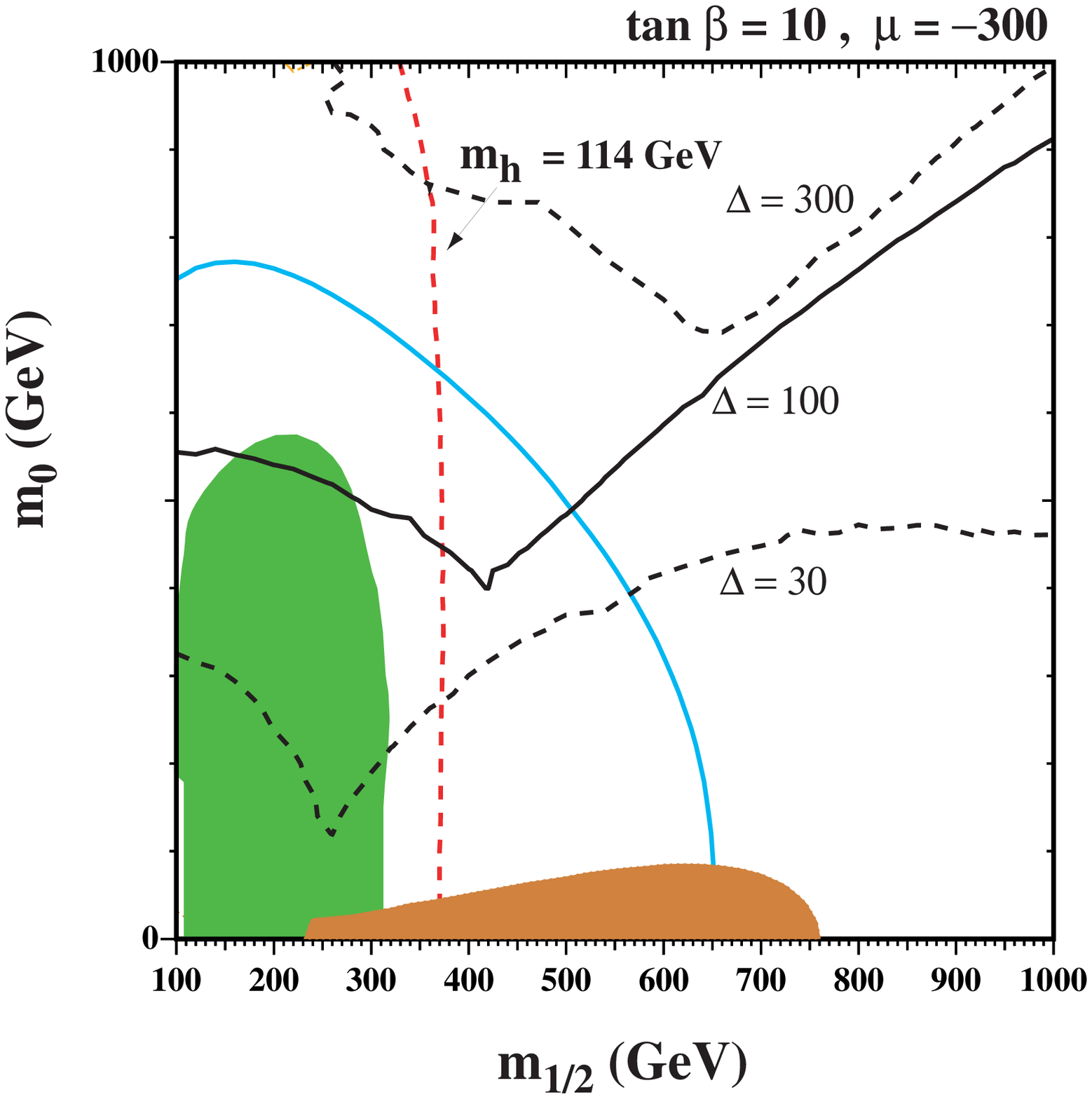,height=3.3in}
%\hspace*{-0.17in}
\epsfig{file=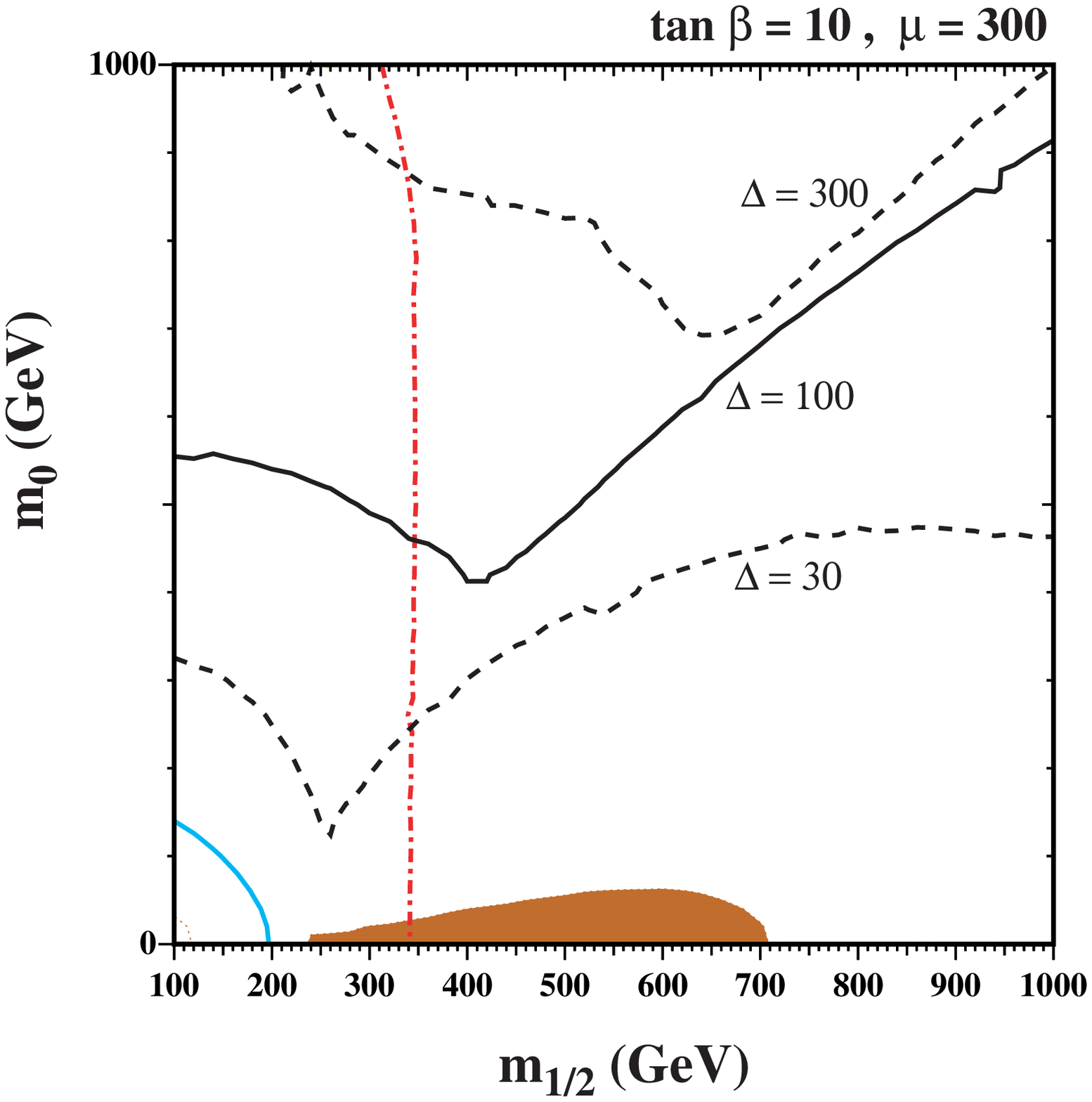,height=3.3in} \hfill
\end{minipage}
%\vspace*{-3in}
%\hspace*{-.70in}
\hspace*{-.20in}
\begin{minipage}{8in}
%\hskip -1.40in
%\vskip -.75in
\epsfig{file=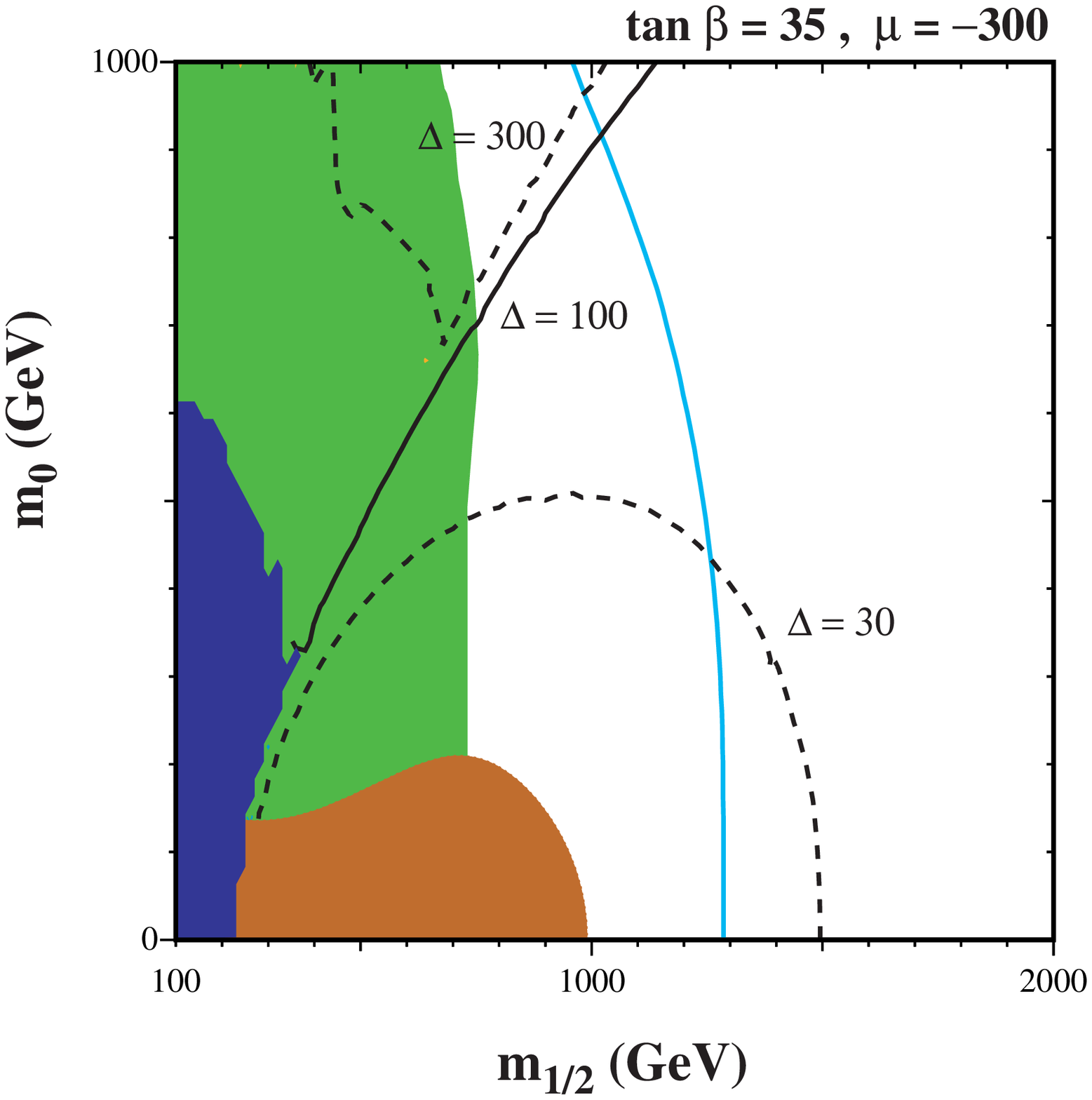,height=3.3in}
\hspace*{-0.10in}
\epsfig{file=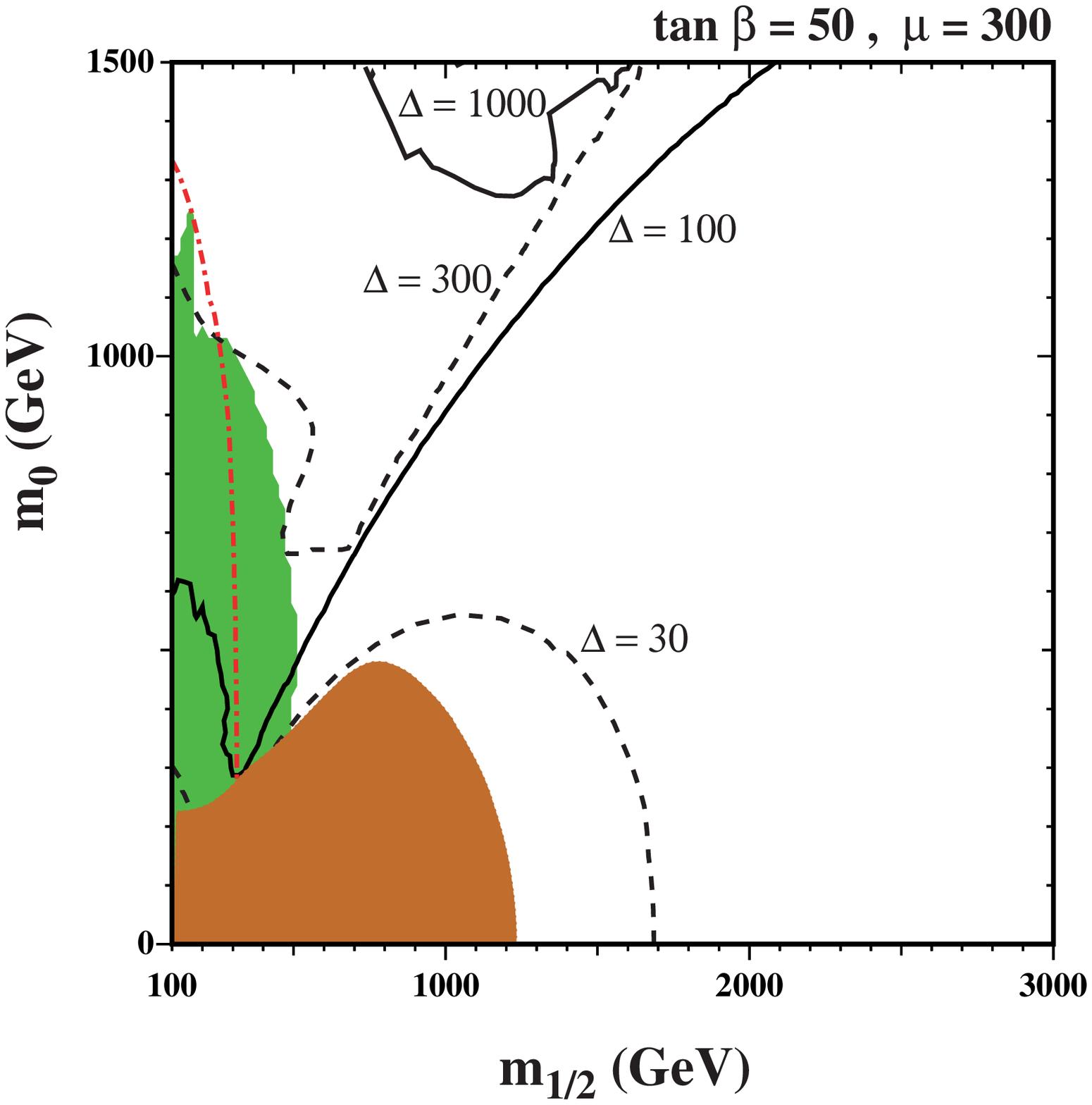,height=3.3in} \hfill
\end{minipage}
%\vskip 2.5in 
\caption{\label{fig:NUFT}
{\it 
Contours of the electroweak fine-tuning parameter $\Delta$
(\ref{thirteen}) for non-universal Higgs masses, in the
$(m_{1/2}, m_0)$ planes for (a) $\tan \beta = 10, \mu = -300$~GeV, (b) $\tan
\beta = 10, \mu = 300$~GeV, (c)
$\tan \beta = 35, \mu = -300$~GeV, and (d) $\tan \beta =
50, \mu = 300$~GeV, all for $m_t = 175$~GeV, $A_0 = 0$ and $m_A
= 1$~TeV. In the dark (brick red) shaded
regions, the LSP is the charged ${\tilde \tau}_1$, so this region is
excluded. In panel (c), the very dark shaded (dark blue) region is
excluded by the electroweak vacuum conditions.
The solid (turquoise) curve shows the  $2 \, \sigma$ $g-2$ bound. }}
\end{figure}  

It is important to note that the relic-density fine-tuning measure
(\ref{twelve}) is distinct from the traditional measure of the fine-tuning
of the electroweak scale~\cite{EENZ}:
\beq
\Delta = \sqrt{\sum_i ~~\Delta_i^{\hspace{0.05in} 2}}\, , \quad \Delta_i \equiv
{\partial \ln
m_W\over \partial \ln a_i}
\label{thirteen}
\eeq
Sample contours of the electroweak fine-tuning measure are shown 
(\ref{thirteen}) are shown in Figs.~\ref{fig:EWFT}.
This electroweak fine tuning is logically different from
the cosmological fine tuning, and values
of $\Delta$ are not necessarily related to values of
$\Delta^\Omega$, as is apparent when comparing the contours in 
Figs.~\ref{fig:overall} and \ref{fig:EWFT}.  Electroweak fine-tuning is 
sometimes used as a 
criterion
for restricting the CMSSM parameters. However, the interpretation of 
$\Delta$ (\ref{thirteen}) is unclear. How large a value of $\Delta$ is
tolerable? Different physicists may well have different pain thresholds.
Moreover, correlations between input parameters may reduce its value in
specific models. 

Note that, the regions allowed by the different constraints can be very
different from those in the CMSSM when we relax some of the CMSSM assumptions,
e.g. the universality between the input Higgs masses and those of the squarks
and sleptons, a subject too broad for complete study in this paper.
As an exercise, we display in Fig.~\ref{fig:NUFT} the electroweak
fine-tuning contours in the Non Universal Higgs Mass model (NUHM), 
where the soft breaking mass terms for the Higgs are not set to equal to $m_0$,
but are derived from the electroweak symmetry breaking condition with two
additional free parameters $m_A$ and $\mu$.

\section{Prospects for Observing Supersymmetry at Accelerators}

\begin{figure}
%\vspace*{-0.75in}
%\hspace*{.40in}
\begin{center}
%\begin{minipage}{8in}
\epsfig{file=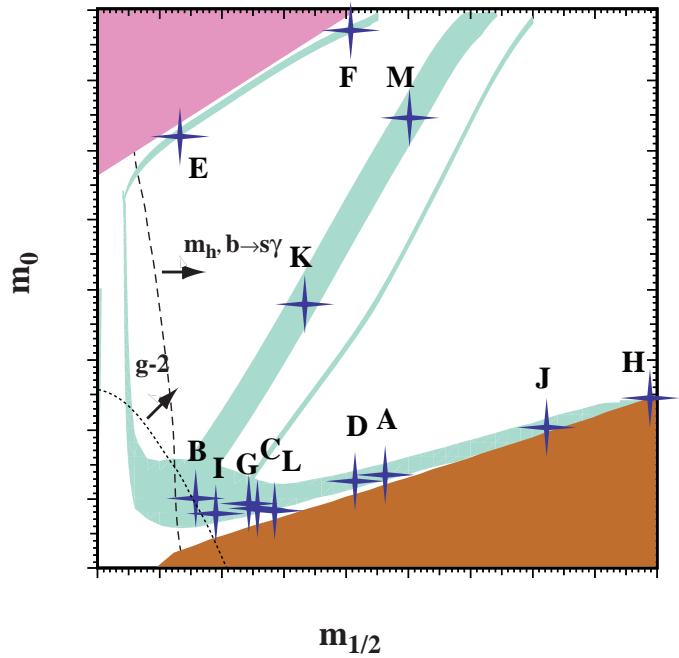,height=3.5in}
\end{center}
%\end{minipage}
\caption[]{\it Schematic overview of the CMSSM benchmark points proposed 
in~\cite{benchmark}.
They were chosen to be compatible with the indicated experimental
constraints, as well as have a relic density in the preferred range
(\ref{ten}). The points are intended to illustrate the range of available
possibilities. The labels correspond to the approximate positions of
the benchmark points in the $(m_{1/2}, m_0)$ plane.}
\label{fig:Bench}
\end{figure}

As an aid to the assessment of the prospects for detecting sparticles at
different accelerators, benchmark sets of supersymmetric parameters have
often been found useful~\cite{benchmarks}, since they provide a focus for
concentrated discussion. A set of proposed post-LEP benchmark scenarios in
the CMSSM~\cite{benchmark} are illustrated schematically in
Fig.~\ref{fig:Bench}. They take into account the direct searches for
sparticles and Higgs bosons, $b\rightarrow s\gamma$ and the preferred
cosmological density range (\ref{ten}). About a half of the proposed
benchmark points are consistent with $g_\mu -2$ at the
$2 \, \sigma$ level, but this was not imposed as an absolute requirement. 

The proposed points were chosen not to provide an `unbiased' statistical
sampling of the CMSSM parameter space, whatever that means in the absence
of a plausible {\it a priori} measure, but rather are intended to
illustrate the different possibilities that are still allowed by the
present constraints~\cite{benchmark}~\footnote{This study is restricted 
to $A = 0$, for which $\stop_1 - \chi$
coannihilation is less important, so this effect has not influenced 
the selection of benchmark points.}. Five of the
chosen points are in the
`bulk' region at small $m_{1/2}$ and $m_0$, four are spread along the
coannihilation `tail' at larger $m_{1/2}$ for various values of
$\tan\beta$, two are in the `focus-point' region at large $m_0$, and two
are in rapid-annihilation `funnels' at large $m_{1/2}$ and $m_0$. The
proposed points range over the allowed values of $\tan\beta$ between 5 and
50.  Most of them have $\mu > 0$, as favoured by $g_\mu - 2$, but there
are two points with $\mu < 0$. All but one point are consistent with the
revised value of $a_\mu$. 

Various derived quantities in these supersymmetric benchmark scenarios,
including the relic density, $g_\mu - 2, b \rightarrow s\gamma$,
electroweak fine-tuning $\Delta$ and the relic-density sensitivity
$\Delta^\Omega$, are given in~\cite{benchmark}. These enable the reader to
see at a glance which models would be excluded by which refinement of the
experimental value of $g_\mu - 2$. Likewise, if you find some amount of
fine-tuning uncomfortably large, then you are free to discard the
corresponding models.

\begin{figure}
%\vspace*{-0.75in}
%\hspace*{.40in}
\begin{center}
%\begin{minipage}{8in}
\epsfig{file=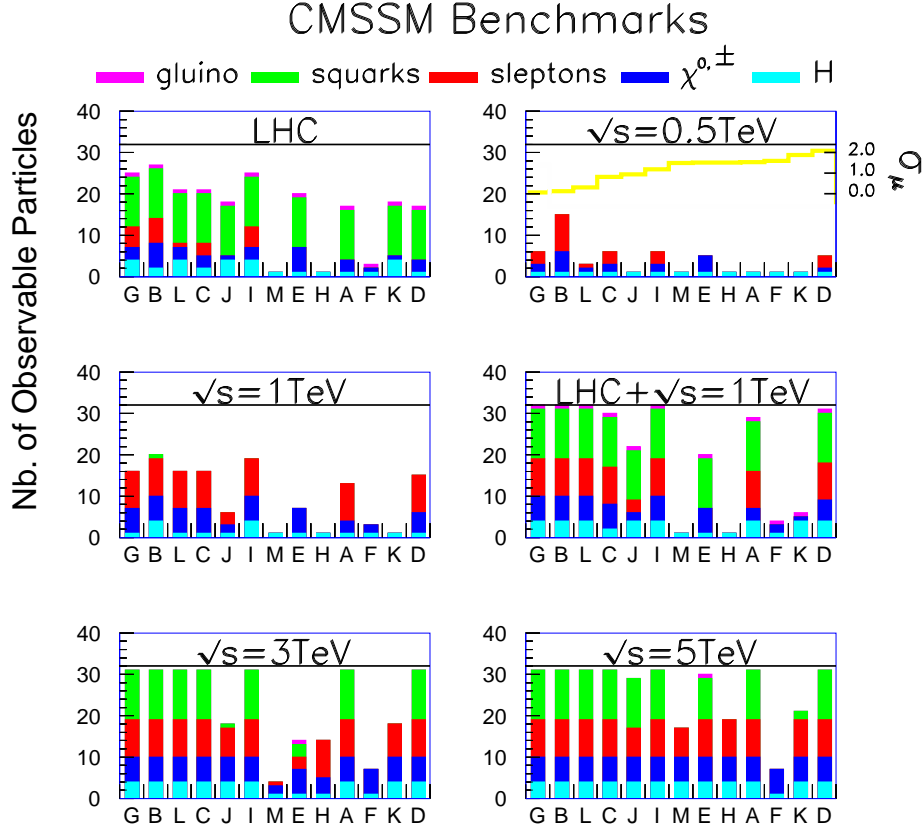,height=5in}
\end{center}
%\end{minipage}
\caption[]{\it 
Summary of the prospective sensitivities of the LHC and
linear
colliders at 
different $\sqrt{s}$ energies to CMSSM particle production 
in the proposed benchmark scenarios G, B, ..., which are
ordered by their distance from the central value of $g_\mu - 2$, as
indicated by the pale (yellow) line in the second panel. We see clearly
the complementarity
between an $e^+ e^-$ collider~\cite{LC,CLIC} (or $\mu^+ 
\mu^-$ collider~\cite{MC}) and the LHC in the TeV
range of energies~\cite{benchmark}, with the former
excelling for non-strongly-interacting particles, and the LHC for
strongly-interacting sparticles and their cascade decays. CLIC \cite{CLIC}
provides unparallelled physics reach for non-strongly-interacting
sparticles, extending beyond the TeV scale. We
recall that mass and coupling measurements at $e^+ e^-$ colliders
are usually much cleaner and more precise than at
hadron-hadron colliders such as the LHC. Note, in particular, that it is
not known how to distinguish the light squark flavours at the LHC. }
\label{fig:Manhattan} 
\end{figure}

The LHC collaborations have analyzed their reach for sparticle detection
in both generic studies and specific benchmark scenarios proposed
previously~\cite{susyLHC}. Based on these studies,
Fig.~\ref{fig:Manhattan} displays estimates of how many different
sparticles may be seen at the LHC in each of the newly-proposed benchmark
scenarios~\cite{benchmark}. The lightest Higgs boson is always found, and
squarks and gluinos are usually found, though there are some scenarios
where no sparticles are found at the LHC. The LHC often misses heavier
weakly-interacting sparticles such as charginos, neutralinos, sleptons and
the other Higgs bosons.

It was initially thought that the discovery of supersymmetry at the LHC
was `guaranteed' if the BNL measurement $g_\mu -2$ was within $2 \, 
\sigma$
of the true value, but this is no longer the case with the new sign of the
pole contributions to light-by-light scattering. This is the case, in 
particular, because arbitrarily large values of $m_{1/2}$ and $m_0$ are 
now compatible with the data at the $2 \, \sigma$ level \cite{bench2}.

The physics capabilities of linear $e^+e^-$ colliders are amply documented
in various design studies~\cite{LC}. Not only is the lightest MSSM Higgs
boson observed, but its major decay modes can be measured with high
accuracy. Moreover, if sparticles are light enough to be produced, their
masses and other properties can be measured very precisely, enabling
models of supersymmetry breaking to be tested~\cite{Zerwas}.

As seen in Fig.~\ref{fig:Manhattan}, the sparticles visible at an $e^+e^-$
collider largely complement those visible at the
LHC~\cite{benchmark,bench2}. In most of benchmark scenarios proposed, a
1-TeV linear collider would be able to discover and measure precisely
several weakly-interacting sparticles that are invisible or difficult to
detect at the LHC. However, there are some benchmark scenarios where the
linear collider (as well as the LHC) fails to discover supersymmetry.
Only a linear collider with a higher centre-of-mass energy appears sure
to cover all the allowed CMSSM parameter space, as seen in the lower
panels of Fig.~\ref{fig:Manhattan}, which illustrate the physics reach of
a higher-energy lepton collider, such as CLIC~\cite{CLIC} or a multi-TeV
muon collider~\cite{MC}.

\section{Prospects for Other Experiments}

\subsection{Detection of Cold Dark Matter}

\begin{figure}
\vskip 0.75in
\vspace*{-0.75in}
\hspace*{-.40in}
\begin{minipage}{8in}
\epsfig{file=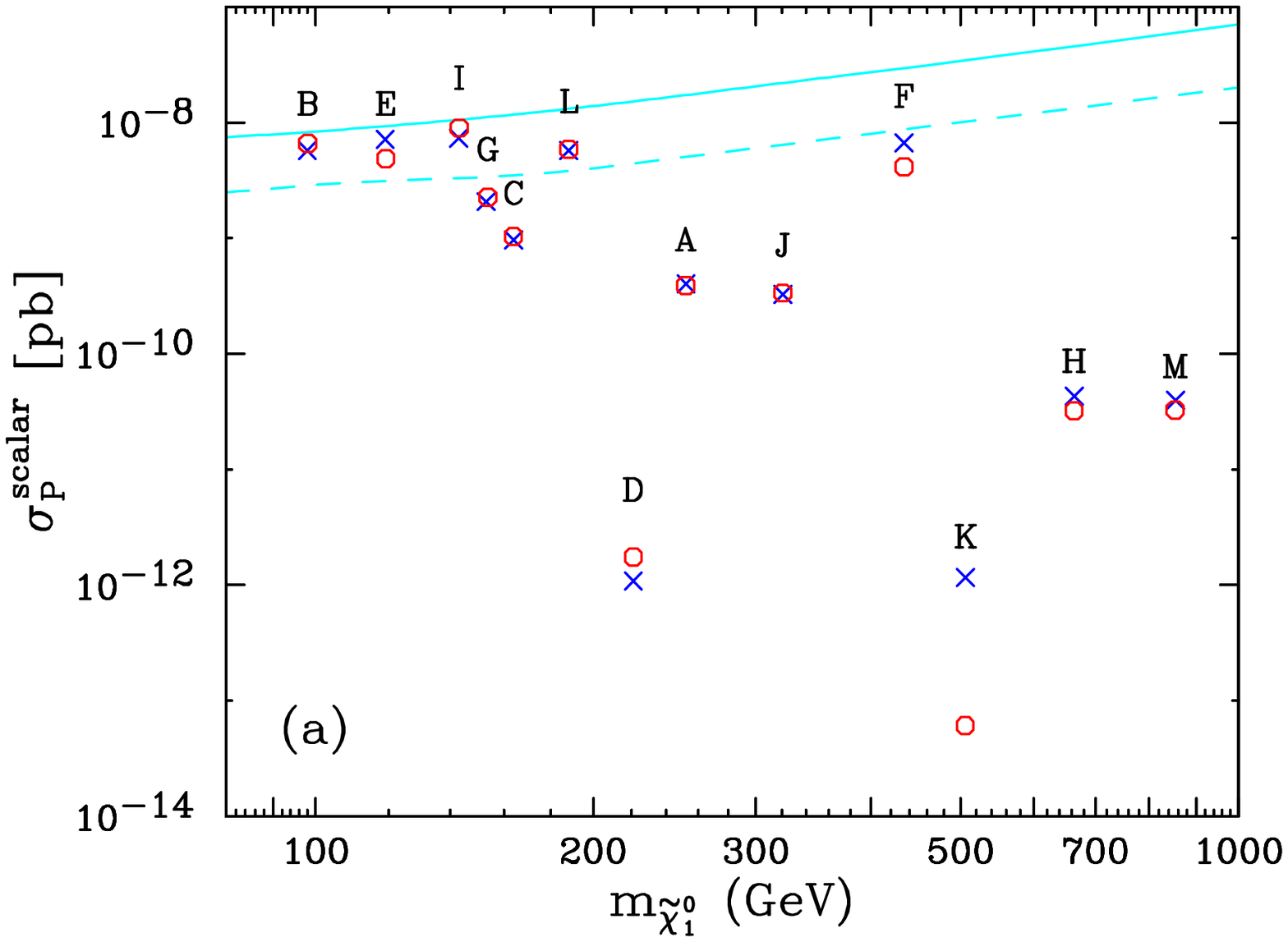,height=2.5in}
\hspace*{-0.17in}
\epsfig{file=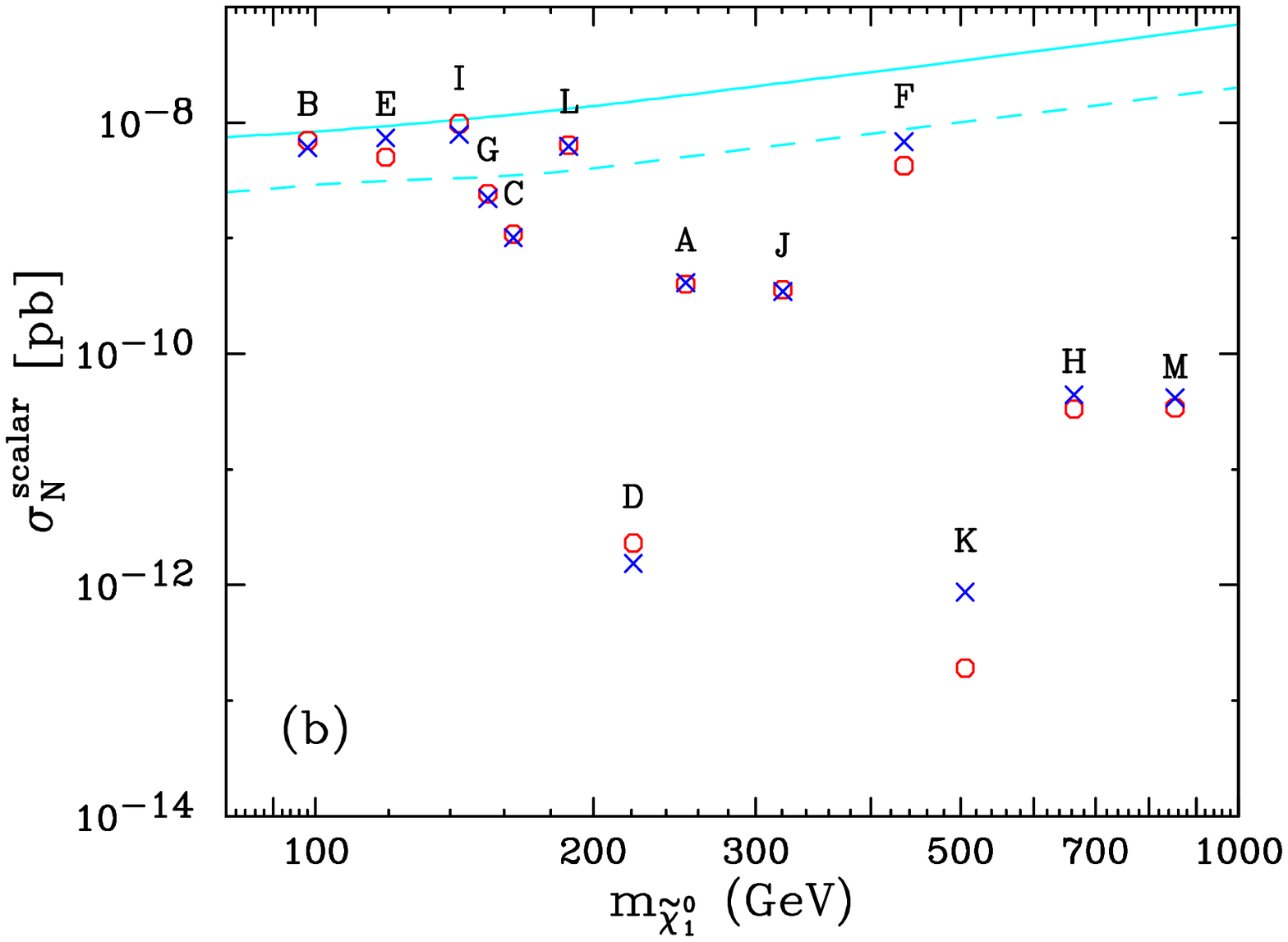,height=2.5in} \hfill
\end{minipage}
\caption[]{\it Elastic spin-independent scattering  
of supersymmetric relics on (a) protons and (b) neutrons calculated in 
benchmark scenarios~\cite{EFFMO}, compared with the 
projected sensitivities for CDMS
II~\cite{Schnee:1998gf} and CRESST~\cite{Bravin:1999fc} (solid) and
GENIUS~\cite{GENIUS} (dashed).
The predictions of our code (blue
crosses) and {\tt Neutdriver}\cite{neut} (red circles) for
neutralino-nucleon scattering are compared.
The labels A, B, ...,L correspond to the benchmark points as shown in 
Fig.~\protect\ref{fig:Bench}.}
\label{fig:DM}
\end{figure}

Fig.~\ref{fig:DM} shows rates for the elastic spin-independent scattering
of supersymmetric relics~\cite{EFFMO}, including the
projected sensitivities for CDMS
II~\cite{Schnee:1998gf} and CRESST~\cite{Bravin:1999fc} (solid) and
GENIUS~\cite{GENIUS} (dashed).
Also shown are the cross sections 
calculated in the proposed benchmark scenarios discussed in the previous
section, which are considerably below the DAMA \cite{DAMA} range
($10^{-5} - 10^{-6}$~pb), but may be within reach of future projects.
Indirect searches for supersymmetric dark matter via the products of
annihilations in the galactic halo or inside the Sun also have prospects
in some of the benchmark scenarios~\cite{EFFMO}.

\subsection{Proton Decay}

This could be within reach, with $\tau (p\rightarrow e^+\pi^0)$ via a
dimension-six operator possibly $\sim 10^{35}y$ if $m_{GUT} \sim 10^{16}$
GeV as expected in a minimal supersymmetric GUT. Such a model also
suggests that $\tau (p\rightarrow \bar\nu K^+) < 10^{32} y$ via
dimension-five operators~\cite{dim5}, unless measures are taken to
suppress them~\cite{fsu5}. This provides motivation for a next-generation
megaton experiment that could detect proton decay as well as explore new
horizons in neutrino physics~\cite{UNO}.

\section{Conclusions}

We have compiled in this short review the various experimental constraints 
on the MSSM, particularly in its constrained CMSSM version. These have 
been compared and combined with the cosmological constraint on the relic 
dark matter density. As we have shown, there is good overall compatibility 
between these various constraints. To exemplify the possible types of
supersymmetric phenomenology compatible with all these constraints, a set 
of benchmark scenarios have been proposed.

We have discussed the fine-tuning of parameters required for supersymmetry 
to have escaped detection so far. There are regions of parameter space 
where the neutralino relic density is rather sensitive to the exact values 
of the input parameters, and to the details of the calculations based on 
them. However, there are generic domains of parameter space where 
supersymmetric dark matter is quite natural. The fine-tuning price of the 
electroweak supersymmetry-breaking scale has been increased by the 
experimental constraints due to LEP, in particular, but its significance 
remains debatable.

As illustrated by these benchmark scenarios, future colliders such as the
LHC and a TeV-scale linear $e^+e^-$ collider have good prospects of
discovering supersymmetry and making detailed measurements. There are also
significant prospects for discovering supersymmetry via searches for cold
dark matter particles, and searches for proton decay also have interesting
prospects in supersymmetric GUT models.

One may be disappointed that supersymmetry has not already been
discovered, but one should not be disheartened. Most of the energy range
where supersymmetry is expected to appear has yet to be explored. Future
accelerators will be able to complete the search for supersymmetry, but
they may be scooped by non-accelerator experiments. In a few years' time,
we expect to be writing about the discovery of supersymmetry, not just
constraints on its existence.

\end{document}